\documentclass[aps,pra,superscriptaddress,twocolumn,floatfix,longbibliography]{revtex4-2}
\usepackage[utf8]{inputenc}
\usepackage[T1]{fontenc}
\usepackage{graphicx,color}
\usepackage[utf8]{inputenc}
\usepackage{amsmath,amssymb,bm,amsfonts,dsfont,mathrsfs,amsthm}
\usepackage{braket}
\usepackage{comment}
\usepackage[colorlinks=true,linkcolor=blue,citecolor=blue,urlcolor=blue]{hyperref}
\usepackage{soul}
\usepackage{tikz}
\usetikzlibrary{shapes, arrows, positioning, calc, backgrounds}

\newcommand{\orcid}[1]{\href{https://orcid.org/#1}{\textcolor[HTML]{A6CE39}{\aiOrcid}}}

\usepackage[caption=false]{subfig}

\begin{document}

\author{Qingyu Li}
\affiliation{Institute of Fundamental and Frontier Sciences, University of Electronic Science and Technology of China, Chengdu 611731, China}
\affiliation{Key Laboratory of Quantum Physics and Photonic Quantum Information, Ministry of Education, University of Electronic Science and Technology of China, Chengdu 611731, China}

\author{Chiranjib Mukhopadhyay}
\affiliation{Institute of Fundamental and Frontier Sciences, University of Electronic Science and Technology of China, Chengdu 611731, China}
\affiliation{Key Laboratory of Quantum Physics and Photonic Quantum Information, Ministry of Education, University of Electronic Science and Technology of China, Chengdu 611731, China}

\author{Ludovico Minati}
\affiliation{School of Life Science and Technology, University of Electronic Science and Technology of China,  Chengdu 611731, China}
\affiliation{Nano Sensing Research Unit, Institute of Innovative Research, Institute of Science Tokyo, 226-8503 Yokohama, Japan}
\affiliation{Center for Mind/Brain Sciences (CIMeC), University of Trento, 38123 Trento, Italy}

\author{Abolfazl Bayat}
\affiliation{Institute of Fundamental and Frontier Sciences, University of Electronic Sciences and Technology of China, Chengdu 611731, China}
\affiliation{Key Laboratory of Quantum Physics and Photonic Quantum Information, Ministry of Education, University of Electronic Science and Technology of China, Chengdu 611731, China}
\affiliation{Shimmer Center, Tianfu Jiangxi Laboratory, Chengdu 641419, China}

\title{Quantum reservoir computing for predicting and characterizing chaotic maps  }

%%%%%%%%%%%%%%%%%%%%%%%%%%%%%%%%%%%%%%%%%%%%%%%%%%%%%%%%%%%%%%%%%%%%%%%%%%%%
%%%%%%%%%%%%%%%%%%%%%%%%%%%%%%%  ABSTRACT  %%%%%%%%%%%%%%%%%%%%%%%%%%%%%%%%%
%%%%%%%%%%%%%%%%%%%%%%%%%%%%%%%%%%%%%%%%%%%%%%%%%%%%%%%%%%%%%%%%%%%%%%%%%%%%

\begin{abstract}
Quantum reservoir computing has emerged as a promising paradigm for harnessing quantum systems to process temporal data efficiently by bypassing the costly training of gradient-based learning methods. Here, we demonstrate the capability of this approach to predict and characterize chaotic dynamics in discrete nonlinear maps, exemplified through the logistic and H\'enon maps. While achieving excellent predictive accuracy, we also demonstrate the optimization of training hyperparameters of the quantum reservoir based on the properties of the underlying map, thus paving the way for efficient forecasting with other discrete and continuous time-series data.  Using closed-loop prediction of distant future steps, our protocol discriminates between chaotic and non-chaotic phases without prior knowledge of the underlying map or the nature of the time series. Furthermore, the framework exhibits robustness against decoherence when trained in situ and shows insensitivity to reservoir Hamiltonian variations  as well as robustness to finite sampling error. These results highlight quantum reservoir computing as a scalable and noise-resilient tool for modeling complex dynamical systems, with immediate applicability in near-term quantum hardware.
\end{abstract}
\date{\today}
\maketitle
%\tableofcontents

%%%%%%%%%%%%%%%%%%%%%%%%%%%%%%%%%%%%%%%%%%%%%%%%%%%%%%%%%%%%%%%%%%%%%%%%%%%%
%%%%%%%%%%%%%%%%%%%%%%%%%%%%%%%%%%%%%%%%%%%%%%%%%%%%%%%%%%%%%%%%%%%%%%%%%%%%
%%%%%%%%%%%%%%%%%%%%%%%%%%%%%%%%%%%%%%%%%%%%%%%%%%%%%%%%%%%%%%%%%%%%%%%%%%%%
\section{Introduction} 
Quantum algorithms with proven quantum advantage, such as prime number factorization~\cite{shorAlgorithmsQuantumComputation1994,vandersypenExperimentalRealizationShors2001} and search~\cite{groverFastQuantumMechanical1996,groverQuantumComputersCan1998}, are not easily implementable in near-term Noisy Intermediate Scale Quantum (NISQ)  computers~\cite{preskillQuantumComputingNISQ2018}. 
Thus, the quest for developing experiment-friendly heuristic quantum algorithms~\cite{Banchi_2024,Khosrojerdi_2025, zimboras2025myths}and identifying new applications is especially important in the current NISQ era. 
Variational quantum algorithms~\cite{Peruzzo_2014,bhartiNoisyIntermediatescaleQuantum2022, Yuan_2019,Tilly_2022,cerezoVariationalQuantumAlgorithms2021,Li_2023} are one class of such algorithms which have attracted a lot of attention recently as a natural quantum extension of neural network-based learning models.
However, they suffer from the same flaws as their classical counterparts - barren plateaus inhibiting training~\cite{mccleanBarrenPlateausQuantum2018, Wang2021noise,Ragone2024lie}, as well as limitations in processing temporal correlations ~\cite{jones2024benchmarkingquantummodelstimeseries}. Recently, quantum systems have been shown to achieve efficiencies beyond classical models for analyzing time series data~\cite{elliottQuantumAdaptiveAgents2022}. In addition, inspired by advancements in classical echo state networks~\cite{jaeger2004harnessing,maass2002real}, quantum reservoir computing (QRC) has emerged as an alternative paradigm of heuristic quantum algorithms, with the benefit of easier training and efficient capture of temporal correlations~\cite{fujii2017harnessing,Fujii2021,innocentiPotentialLimitationsQuantum2023a}. QRC has already found applications for time series analysis~\cite{fujii2017harnessing,chenLearningNonlinearInput2019,xiaConfiguredQuantumReservoir2023,li2025quantumreservoircomputingrealized}, entanglement detection~\cite{ghoshQuantumReservoirProcessing2019,innocentiPotentialLimitationsQuantum2023a}, quantum tomography~\cite{angelatosReservoirComputingApproach2021,krisnandaTomographicCompletenessRobustness2023}, quantum estimation~\cite{krisnandaPhaseMeasurementStandard2022,liEstimatingManyProperties2024} and quantum state preparation~\cite{supranoExperimentalPropertyReconstruction2024,ghoshQuantumNeuromorphicPlatform2019} among others. QRC has also been proposed for various quantum platforms, such as photonic~\cite{ghoshQuantumReservoirProcessing2019,supranoExperimentalPropertyReconstruction2024,ZhuPractical2025}, coherently coupled quantum oscillators~\cite{dudasQuantumReservoirComputing2023}, neutral atomic Rydberg arrays~\cite{bravoQuantumReservoirComputing2022}, nuclear spin-based reservoirs~\cite{negoroMachineLearningControllable2018}, and superconducting qubit platforms~\cite{chenTemporalInformationProcessing2020}. 
While the prediction capacity of QRC has been demonstrated for time series~\cite{fujii2017harnessing,tranLearningTemporalQuantum2021,zhangLearningHamiltonianDynamics2021,suzukiNaturalQuantumReservoir2022,mujalTimeseriesQuantumReservoir2023,kobayashiFeedbackDrivenQuantumReservoir2024,salatino2025forecastinglowdimensionalturbulencemultidimensional,delorenzis2025harnessing,li2025quantumreservoircomputingrealized,hou2025highaccuracy}, its full potential for the prediction and characterization of signals with inherent nonlinearity, in particular generated by chaotic behavior, is less explored.

Classical discrete non-linear systems, which describe various situations, exhibit rich dynamical behaviors~\cite{maySimpleMathematicalModels1976}. These include sinks or stable fixed points, periodic orbits, limit cycles, and, strikingly, chaotic behavior where slight changes in initial conditions lead to divergent outcomes~\cite{strogatzNonlinearDynamicsChaos2019}. This is formalized through the notion of Lyapunov exponents, which characterizes the rate of separation of infinitesimally close trajectories.  
Formally, if two series $\{x_t\}$ and $\{x'_t\}$ emanating from the same dynamical system with slight differences in the values of the control parameters are considered with initial separation $\delta_0{=}x_0{-}x'_0$, the largest Lyapunov Exponent (LLE) $\lambda^*$ is defined as $\lambda^*{=}\lim_{t\to\infty,|\delta(0)|\to 0}(1/t)\ln{(|\delta(t)/\delta(0)|)}$, where $\delta_t{=}x_t{-}x'_t$. 
If $\lambda^* {<} 0$, the perturbations in the initial conditions decay exponentially, indicating that the system is stable. If $\lambda^* {=} 0$, the system exhibits neutral stability commonly associated with periodic or quasi-periodic dynamics. In contrast, if $\lambda^*{>}0$, the trajectories originating from infinitely close initial conditions diverge exponentially over time, a hallmark of chaotic behavior. This makes the prediction of future outcomes of time series generated from chaotic maps a truly challenging task. Thus, several questions arise: (i) Can quantum machine learning algorithms, such as QRC, succeed in forecasting? and (ii) Can QRC setups be used for characterizing the underlying features, such as nonlinearity and memory?

\begin{figure}
    \includegraphics[width = \linewidth]{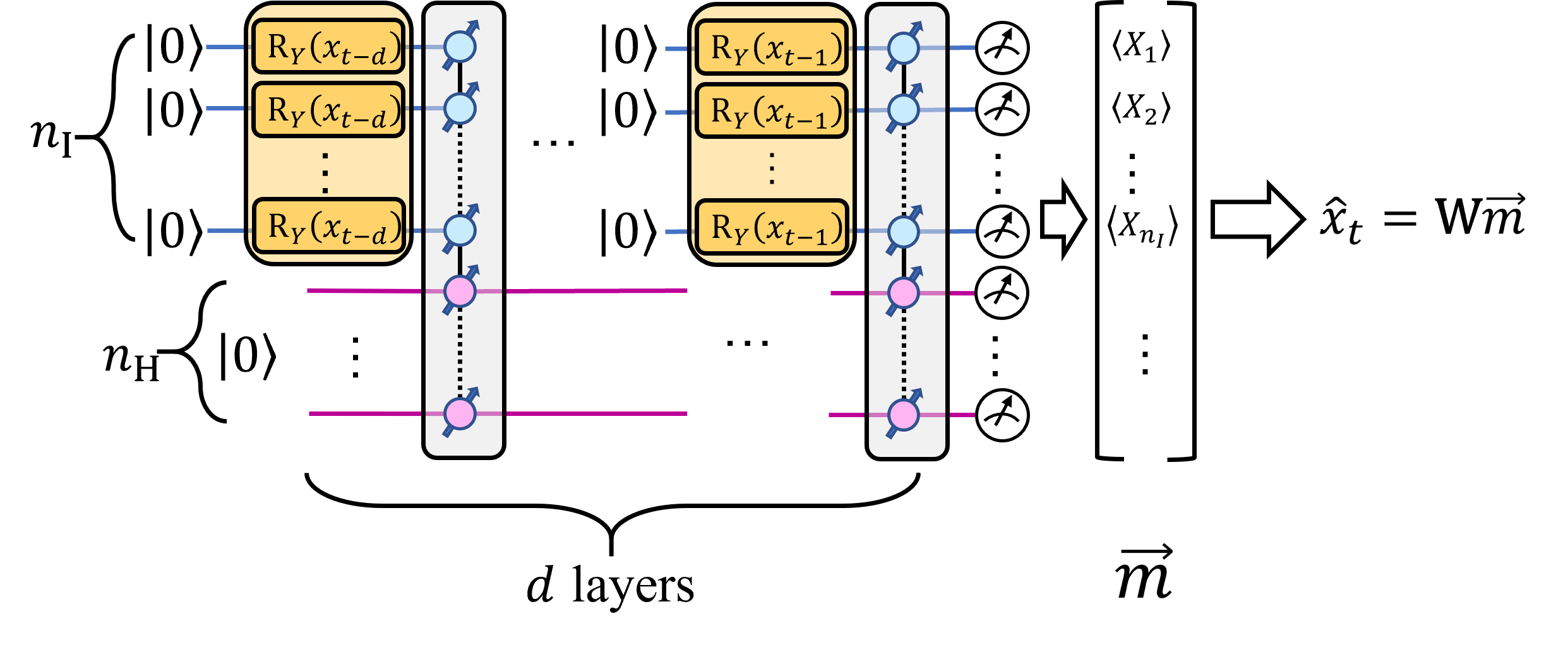}
    \caption{Schematic of QRC architecture. Quantum reservoir comprises two subsystems: (i) input qubits (blue) for encoding input variables through Pauli $Y$-rotation $R_Y$, and (ii) hidden qubits (pink) storing historical information to predict future outcomes. This process begins at time step $t{-}d$ and continues until $t{-}1$, such that the total number of quantum layers is $d$.  After the evolutions of $d$ sequential layers, quantum measurements along Pauli $X$-directions are implement on all the input and hidden qubits.}
    \label{fig1:schematic}
\end{figure}

In this work, we answer these questions in the affirmative by demonstrating that discrete nonlinear dynamical maps can indeed be learned by quantum reservoir computing, leading to accurate forecasts even in the presence of fully developed chaotic dynamics. 
%we evaluate its prediction performance, quantum resource requirements, robustness to noise, and the feasibility of using randomly initialized quantum reservoirs. In this study, we%
Specifically, we demonstrate the success of a quantum reservoir consisting of linearly connected XY chains in predicting the behavior of two canonical nonlinear dynamical maps, viz., the logistic map~\cite{maySimpleMathematicalModels1976} and the H\'enon map~\cite{henonTwodimensionalMappingStrange1976}. 
Analyzing the predictive success based on resource consumption strategies such as increasing the number of quantum layers and different encoding repetitions yields valuable insights towards efficient generalization of our approach to arbitrary nonlinear systems. We also perform a detailed analysis of the closed-loop strategy for long-horizon predictions possible with our approach.  
Furthermore, we demonstrate the robustness of our approach through assessing the impact of quantum decoherence noise, random choice of reservoir Hamiltonians, and finite sampling errors on the prediction performance.
\section{Chaotic maps} 
We consider two canonical non-linear dynamical maps, viz. the logistic and the H\'enon maps, which have one and two control parameters, respectively, for generating time series. 
The goal is to use a quantum reservoir algorithm for predicting and characterizing such time series. The logistic map is the one‑dimensional recurrence relation 
\begin{equation}
    x_{t} = rx_{t-1}(1{-}x_{t-1}), 
    \label{eq:logistic}
\end{equation}
\noindent where $r$ is the control parameter generally taken between $r {\in} [0,4]$, and $x_t {\in} [0,1]$. The map stems from Verhulst's logistic growth law of the 19th century, subsequently finding application as a discrete model of population dynamics ~\cite{maySimpleMathematicalModels1976}, and is now widely popular as a pedagogical minimal model to demonstrate chaotic properties, as varying a single parameter $r$ reveals fixed points, period-doubling cascades, windows of periodicity, and fully chaotic regimes ~\cite{strogatzNonlinearDynamicsChaos2019}. 

The H\'enon map is a two‑dimensional discrete map originally introduced to study strange attractors in dissipative systems  ~\cite{henonTwodimensionalMappingStrange1976}. It is given by the two-dimensional recurrence relation 

\begin{equation}
    x_t = 1 {-} ax_{t-1}^2 {+} y_{t-1} \quad; \quad y_t = bx_{t-1} \quad.
    \label{eq:henon}
\end{equation}

\noindent The H\'enon map provides a minimal, analytically accessible model of deterministic chaos in low dimensions.  With $a {=}1.4$ and $b {=} 0.3$, it produces the famous H\'enon attractor. The map represents the stretch‑and‑fold mechanism of chaos in compact polynomial form. Each iteration stretches one direction, folds the trajectories back and contracts the area by $|b|$, creating a fractal invariant set with sensitive dependence on the initial conditions \cite{benedicks1991dynamics}.

\section{Quantum reservoir computing}
In the quantum reservoir computing (QRC) framework for classical time-series forecasting \cite{fujii2017harnessing,ghosh2019quantum,Fujii2021,mujal2021opportunities,mujal2023time,garcia2024quantum}, qubits are encoded with the time-series data at various time steps, and then allowed to sequentially interact with a quantum reservoir, whose Hamiltonian remains fixed throughout the training and prediction processes. This sequential interaction scrambles the encoded information inside the reservoir, which is then decoded through a final set of measurements enabling us to make a prediction for future time-steps. Unlike the paradigm of variational quantum algorithms, where one seeks to iteratively update all the parameters of a parametrized quantum circuit to optimize a given loss function, the QRC approach employs training only once, exclusively at the final readout stage. This mirrors classical reservoir computing, with the  difference that the reservoir is implemented through a quantum system rather than a classical one.

\section{QRC protocol for chaotic map prediction }
Let us now elaborate our implementation of QRC as depicted in Fig.~\ref{fig1:schematic} step-by-step. The reservoir collectively consists of two components - (i) qubits encoding information about the dynamical map (blue qubits in Fig.~\ref{fig1:schematic}), which are replaced after each layer with fresh qubits encoding updated information; and (ii) so-called hidden qubits (pink qubits in Fig.~\ref{fig1:schematic}) which are not directly manipulated until the very end, when at the output layer, all qubits in the reservoir are measured locally. The QRC architecture is as follows. 

\emph{(i) Reservoir engineering--} Our reservoir is a linearly connected transverse XY chain with Hamiltonian given by 
\begin{equation}
    H{=}J\sum_{j=1}^{N}(X_jX_{j+1}{+}Y_jY_{j+1}){+}\sum_j h_jZ_j,
\end{equation}
where $\{X_j,Y_j,Z_j\}$ are respectively the Pauli operators acting on the $j$-th qubit. The exchange couplings $J$ are fixed to unity in this study, and the magnetic field strengths $\{h_i\}$ are randomly sampled from within the ordered phase $h_i {\in} [0,1]$.

\emph{(ii) Encoding--} 
Each input data $\vec{x}_i$ is a segment of a time series $i$ of the form $\vec{x}_i{=} (x^{(i)}_{t-d}, x^{(i)}_{t-d+1},\cdots,x^{(i)}_{t-1})$, generated from a chaotic map, in which the last $d$ values are used to predict the next step $x^{(i)}_{t}$.  For encoding the data, each data point of $\vec{x}_i$ is fed, oldest first, into the reservoir through $d$ sequential layers, where each layer consists of one elements of $\vec{x}_i$ encoding into qubits through a Pauli Y-rotation. 
To introduce higher-order terms~\cite{mitaraiQuantumCircuitLearning2018,schuldEffectDataEncoding2021}, $n_{\textrm{rep}}$~repetitions of the encoded qubit may be deployed at each layer. For example, when $n_{\textrm{rep}}=1$, the input state is $R_Y(x)|0\rangle=\cos(x)|0\rangle+\sin(x)|1\rangle$. However, when we increase $n_{\textrm{rep}}=2$, the input state is $R_Y(x)|0\rangle \otimes R_Y(x)|0\rangle=\cos^2(x)|00\rangle+\sin(x)\cos(x)(|01\rangle+|10\rangle)+\sin^2(x)|11\rangle$. Thus, progressively higher order nonlinear terms are injected into the input state by increasing $n_{\textrm{rep}}$.  
After each layer, the reservoir evolves under Hamiltonian $H$ for a time $\tau$, following which the original encoding qubits are discarded (i.e. traced out) and replaced by fresh qubits encoding the next elements of $\vec{x}_i$ and the state of hidden qubits $\rho_\textrm{H}$ is maintained to the next layer. This process continues until the information at time step $t{-}1$ is fed into the quantum reservoir.

\emph{(iii) Readout--} After the final round of evolution following the encoding of $x^{(i)}_{t-1}$, all input and hidden qubits (the numbers are denoted as $n_\textrm{I}$ and $n_\textrm{H}$, respectively) in the reservoir are locally measured along Pauli X-direction to yield a vector of expectation values $\vec{m}_i{=}[\langle X_1\rangle, \langle X_2\rangle, \cdots, \langle X_{n_\textrm{I}+n_\textrm{H}}\rangle]^T $. 

\emph{(iv)~Training and Prediction--}   
The core of QRC is its simple prediction method which is based on linear regression. In this approach,  $x^{(i)}_{t}$ is estimated through a linear map like $x^{(i)}_{t}{=}\textbf{W}\vec{m}_i$ in which the weight matrix $\textbf{W}$ has to be trained. This training is repeated through rolling windows over $s$ different training data sets $\{\vec{x}_1, \vec{x}_2, \cdots, \vec{x}_s\}$, each yielding a different measurement outcome $\vec{m}_i$. By making a matrix of such measurement outcomes $\textbf{M}{=}[\vec{m}_1, \vec{m}_2, \cdots, \vec{m}_s]$ and considering the true values of the time series $\textbf{Y}{=}[x^{(1)}_t,x^{(2)}_t,\cdots,x^{(s)}_t]$ one can train the weight matrix $\textbf{W}$ by minimizing the mean square error between the predicted value of $\textbf{W}\vec{m}_i$ and the corresponding true value $x^{(i)}_t$, namely $\text{MSE}{=}1/s\sum_i^s(x_t^{(i)}-\textbf{W}\vec{m}_i)^2$. 
%\begin{equation}
%    \text{MSE}=\frac{1}{s}\sum_i^s(x_t^{(i)}-\textbf{W}\vec{m}_i)^2.
%\end{equation}
This results in the optimal weight matrix $\textbf{W}$ as
\begin{equation}
    \textbf{W}^*=\textbf{Y}\textbf{M}^\top(\textbf{M} \textbf{M}^\top+\epsilon I)^{-1},
    %\textbf{W}^*{=}\textbf{Y}\textbf{M}^{+}
\end{equation}
%where $\textbf{M}^{+}$ is the Moor-Penrose pseudo-inverse of the outcome matrix. 
where $I$ is the identity matrix and $\epsilon$ is a small value (we take it as $10^{-8}$ in this study) to ensure the robustness of the numerical calculation for matrix inversion.
Hence, for predicting the unseen data point $x_t$ based on the $d$ past known values $(x_{t-d},x_{t-d+1},\cdots,x_{t-1})$, these values are similarly fed into the reservoir layer-by-layer to yield the final outcome vector $\vec{m}$. The prediction $\hat{x}_t$ for $x_t$ is then given by $\hat{x}_t{=}\textbf{W}^*\vec{m}$.

\section{Predicting chaotic time series} 
We are now in a position to illustrate its performance for forecasting of the logistic and H\'enon maps mentioned before. We restrict ourselves to time series generated with $r{\in}[0,4]$ for the logistic map and $a{\in}[1,1.4]$ for the H\'enon map. For the training set, for any given control parameter $r$ or $a$, we consider $100$ different time series with $20$ elements each, namely $\{x^{(i)}_t\}_{t=1}^{20}$ with index $i$ going from $1$ to $100$ denoting each time series. Each time series is generated by a different random initial element $x^{(i)}_1$ and at the end we normalize all the time series with the largest element among all the $100$ time series to ensure that all the elements remain between $0$ and $1$. For the test data, we use the same approach for generating $10$ unseen time series with new initializations, each containing $200$ time steps.

\begin{figure}
    \includegraphics[width = \linewidth]{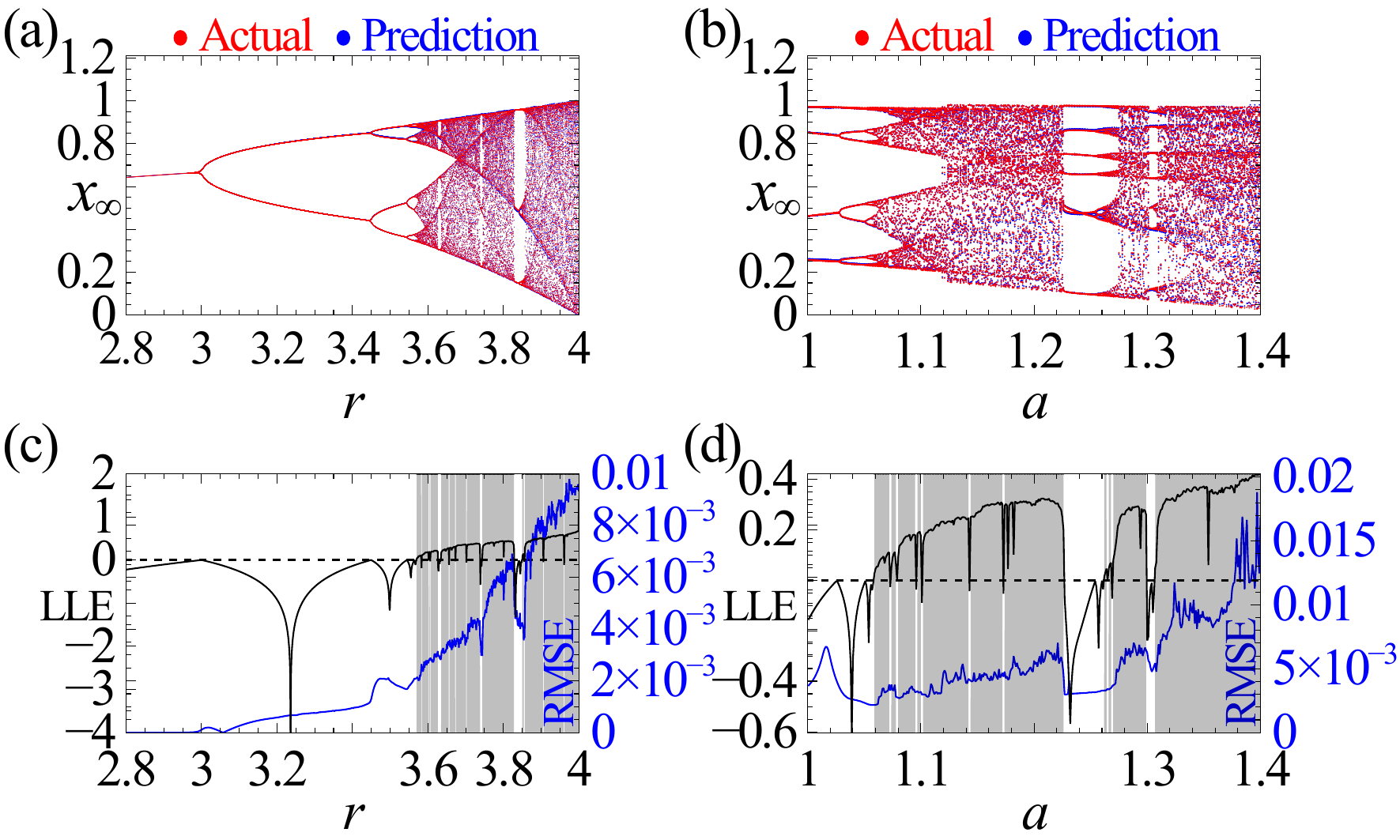}
    \caption{The bifurcation diagram of the logistic map~(a) and the H\'enon map~(b). The largest Lyapunov exponent~(dark line) of the logistic map~(c) and the henon map~(d). The regions with a white background indicate LLE values less than 0, while the grey background highlights regions where the LLE is greater than 0. Additionally, the RMSE of predicting the bifurcation diagrams have shown as the blue lines in (c) and (d).}
    \label{fig2:prediction}
\end{figure}

For the logistic map, we consider a quantum reservoir with $d{=}2$ layers and $n_\textrm{H}{=}4$ hidden qubits. Since the logistic map only depends on one variable, the number of input qubits will be equal to the number of repetitions, namely $n_\textrm{I}{=}n_{\textrm{rep}}$. We keep $n_\textrm{I}{=}n_{\textrm{rep}}{=}2$ for simulating the logistic maps.  For the H\'enon map, we consider a quantum reservoir with $d{=}1$ layer and $n_\textrm{H}{=}3$ hidden qubits. Since this H\'enon map has two input variables, then $n_\textrm{I}{=}2n_\textrm{rep}$. Here, we take $n_\textrm{I}{=}4$ input qubits (i.e. $n_{\textrm{rep}}{=}2$ repetitions for each input data). We use the rolling window method for training.  In Figs.~\ref{fig2:prediction}(a) and (b), for any given control parameters $r$ and $a$, we depict the long time predictions from $t{=}151$ to $t{=}200$ for one of the test time series which shows very accurate prediction (blue points) for our QRC protocol. The bifurcation diagrams in Figs.~\ref{fig2:prediction}(a) and (b) clearly single out the different dynamical regimes of the maps showing remarkable agreement with the corresponding true values (red points).  Note that $x_{\infty}$ in these figures is represented by all the predictions for this test dataset for times between $t={151}$ to $t={200}$. 
As the figure shows, for certain values of the control parameter, the long time behavior shows convergence to either a single value or oscillations between a few discrete values, indicating non-chaotic behavior. On the other hand, for some other values of the control parameters, the long time behavior tends to cover a continuous part of the space, which is an evidence of chaotic behavior.  Interestingly, the predictions by the QRC can precisely identify the islands of periodicity between the chaotic regimes for both maps.
For a more in-depth analysis, from the prediction in all test sets, we plot the corresponding Root Mean Square Error (RMSE) averaged over all test datasets in Figs.~\ref{fig2:prediction}(c) and (d) for the logistic and H\'enon maps respectively, where we also plot the corresponding LLEs for reference. Two results immediately appear -- \emph{(i)} the prediction error is lowered wherever the LLE is negative (sweeping across map parameter ranges -- average of RMSE prediction errors across chaotic regimes vs non-chaotic regimes, which is respectively $5.7\times10^{-5}$ vs $8\times10^{-4}$ for logistic map, and $6.3\times10^{-5}$ vs $3.6\times10^{-5}$ for H\'enon map), and \emph{(ii)} even within the extended regions where the LLE is positive, a very strong correlation (Spearman rank correlation coefficient $r_s{=}0.97$ for the logistic map, and $r_s{=}0.87$ for the H\'enon map) between LLE and prediction error is observed. Together, these results show that the prediction error primarily inherent the degree of unpredictability of the dynamics. \\

%Moreover,  LLE dipping to negative values within the chaotic regime (the so-called islands of stability) is also inevitably identified by lower prediction errors. This is fully consistent with the intuition that divergence of trajectories within the chaotic regime should make prediction harder and vice versa.  \textcolor{black}{Chiranjib will write this with numbers following Ludovico's script}

\section{Characterizing chaotic time series}
\subsection{Open-loop prediction of QRC}
We have already demonstrated that our QRC framework is highly effective in performing single-step predictions. However, this still leaves a major question unaddressed - can one taylor the hyperparameters involved in the QRC implementation protocol to maximize the prediction accuracy for specific nonlinear dynamical maps? Conversely, assuming such an optimization strategy for one member of a specific family of nonlinear dynamical maps is obtained, can this strategy inform the QRC optimization strategy for other members of that family of maps as well?
%$$ although conceptually fundamental, are rarely encountered exactly in real experimental data. Thus, one asks the following question -- does the analysis presented here also serve as a predictive tool to gauge the \sr{degree of nonlinearity} in data?}

\begin{figure}
    \includegraphics[width = \linewidth]{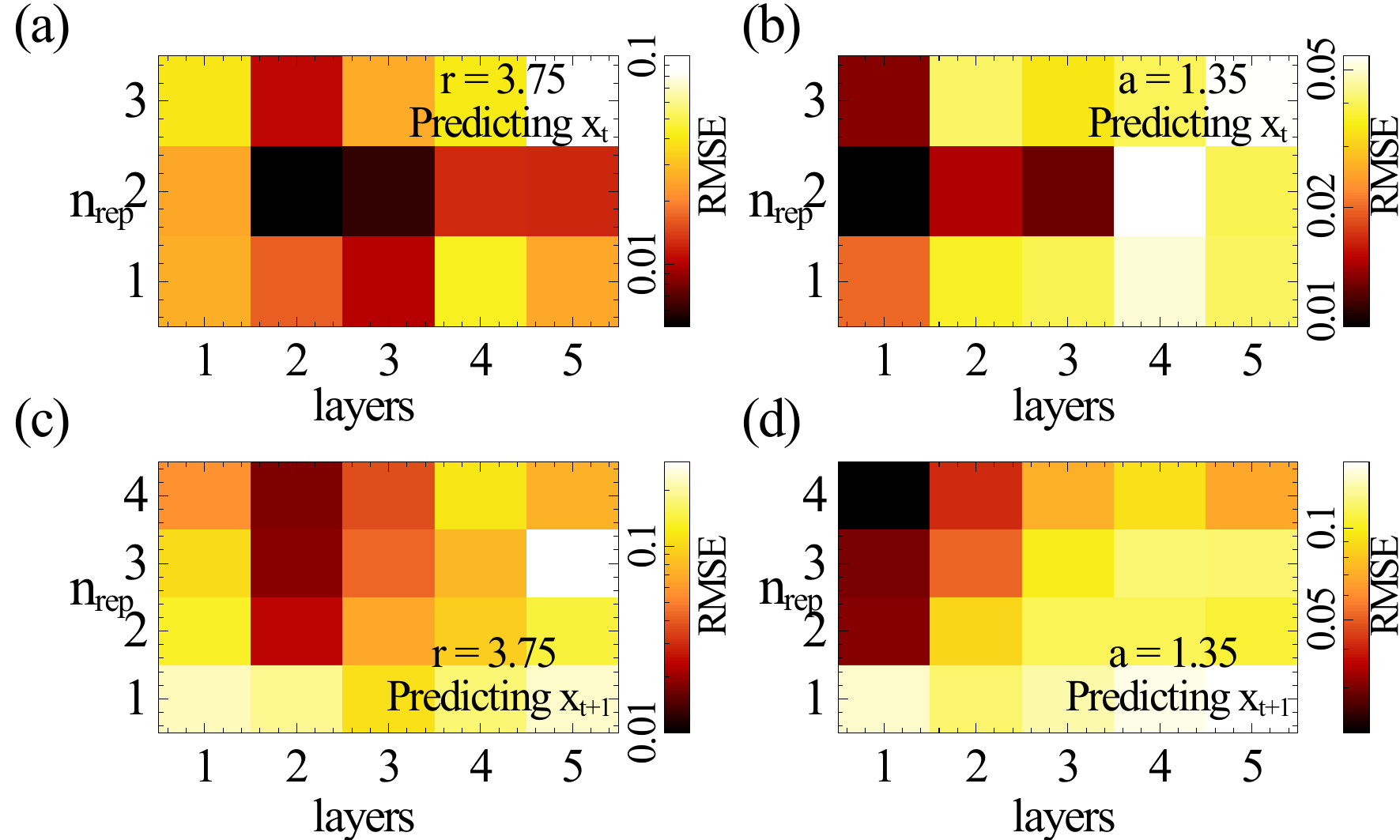}
    \caption{The effect of number of layers and repetitions on QRC for the logistic and H\'enon map for (a) and (b) predicting $x_t$ and (c) and (d) predicting $x_{t+1}$ for $r=3.75$ ($a = 1.35$). Darker hues indicate lower RMSE (better predictive accuracy). Lowest RMSE is for $2$ layers for logistic map ($1$ layer for H\'enon map) and $n_{\textrm{rep}}{=}2$ for predicting $x_t$ and $n_{\textrm{rep}}{=}4$ for predicting $x_{t+1}$ in both cases. }
    \label{fig3:architecture}
\end{figure}

To answer these questions, we confine ourselves to two hyperparameters of QRC, namely the number of repetitions $n_{\textrm{rep}}$ with which encoded qubits are introduced at each layer as well as the number of layers, i.e., circuit depth. Moreover, we consider the following family $\Phi$ of polynomial maps $\Phi : x_{t}{=}\sum_{j} P_{\mu_j}(x_{t-j})$, where $P_{\mu_j}$ is a polynomial of degree $\mu_j$. Both the logistic and the H\'enon maps are members of this family. For example, for the logistic map the only nonzero contribution is coming from $j{=}1$ for which $\mu_1{=}2$. In Fig.~\ref{fig3:architecture}(a) and Fig.~\ref{fig3:architecture}(b), we plot the prediction accuracy as a function of the number of layers as well as repetitions $n_{\textrm{rep}}$ for the logistic and H\'enon maps respectively. The results strongly indicate that increasing the number of quantum layers or repetitions is not monotonic with prediction accuracy. For example, for the logistic map, best predictive accuracy for $x_t$ is achieved with two quantum layers and $n_{\textrm{rep}} {=} 2$, corresponding to the input sequence $\{x_{t-2},x_{t-2},x_{t-1},x_{t-1}\}$. To explain the result, one recalls that the repetitions serve to induce nonlinearity in the otherwise linear quantum circuit. Thus $x_t$ is a nonlinear function of degree 2 of $x_{t-1}$ as per both the logistic and H\'enon maps, therefore the expected optimal number of repetitions  $n_{\textrm{rep}}{=}2$.

From this, we hypothesize that the optimal repetition strategy $n_{\textrm{rep}}^{*}$ for the family of maps $\Phi$ is given by the highest polynomial degree i.e., $\mu_{\max}=\max_j(\mu_j)$ as per the definition of $\Phi$ above. To test this hypothesis, we take the same logistic or H\'enon maps, but now attempt to predict $x_{t+1}$ in terms of only $x_{t-1}$ and elements before. We note that $x_{t+1}$ is a degree-4 function of $x_{t-1}$ in both cases, and are thus also members of the family $\Phi$ in which the only non-zero contribution comes from $j{=1}$ with $\mu_{1}{=}4$. For example, the logistic map is now explicitly given as $x_{t+1} {=} {-}r^3x_{t-1}^4 {+} 2r^3 x_{t-1}^3 {-} r^2 (1+r)x_{t-1}^2 {+} r^2 x_{t-1}$, Thus, we empirically expect $n_{\textrm{rep}}{=}4$ to be optimal for predicting two steps ahead. Indeed, Fig.~\ref{fig3:architecture}(c) and Fig.~\ref{fig3:architecture}(d) explicitly confirm this prediction numerically for both logistic and H\'enon maps and thus support our hypothesis. As a corollary, this hypothesis also entails that quantum resource requirements for predicting chaotic systems over multiple time steps may sharply diverge. One can arrive at this conclusion in two steps. First, one notes that for the problem of predicting $x_{t{+}k-1}$, i.e., $k$-steps in advance, the original map of a member of $\Phi$ with highest polynomial degree $\mu_{\max}$ transforms into another member of $\Phi$ with highest polynomial degree $\mu_{\max}^{k}$. Now, our hypothesis indicates that optimal input repetitions $n_{\textrm{rep}}^*$ must scale as $\mu_{\max}^{k}$, i.e., exponentially with $k$. 
This shows that predicting distant future outcomes of chaotic maps requires exponential resources, in terms of number of qubits.  \\

While the above analysis is extracted for the systems whose underlying maps are known, the same approach can be extended to practical scenarios where the nonlinear degree is unknown. Roughly speaking, in such cases we can offer a bootstrapping mechanism -- inferring the rough nonlinear degree by selecting a known batch of data and checking how many repetitions achieve best prediction on that batch, thus allowing us to form an estimate of the nonlinear degree of the system itself.
%While such exponential scaling explains why predicting time series with inherent nonlinearity is so hard in practice, it nonetheless shows the potential of large scale quantum computers to predict and characterize such signals.  Specifically, the number of repetitions $n_{\textrm{rep}}^{*}$ corresponding to numerically obtained best prediction $\tilde{k}$ steps ahead immediately serves as an estimate to the degree of nonlinearity of the underlying unknown map as $\exp\left[\ln (n_{\textrm{rep}}^{*})/ \tilde{k} \right]$. 

\subsection{Closed-loop prediction of QRC}
In the previous section, we found that performing multi-step-ahead direct prediction can be demanding in terms of quantum resources. For instance, one typically needs a larger number of input repetitions $n_{\mathrm{rep}}$ to induce higher-order nonlinear features, which becomes crucial when forecasting further into the future.
In this section, we investigate how to combine QRC with closed-loop forecasting to progressively predict future values of the logistic and H\'enon maps. 
In a closed-loop setting, the model's own predictions are fed back as inputs for subsequent steps, thereby generating a multi-step trajectory through iterative one-step-ahead prediction.
Specifically, consider a QRC model with $d$ layers. Given an initial input window $(x_{t-d}, x_{t-d+1}, \ldots, x_{t-1})$, our goal is to predict the value at time $t{-}1{+}k$~, with $k$ being the prediction horizon. Rather than training the QRC to output $x_{t-1+k}$ directly (as in the direct multi-step setup for $k{=}2$ discussed earlier), we first use the model to produce a one-step prediction $\hat{x}_{t}$. We then slide the input window forward to $(x_{t-d+1}, \ldots, x_{t-1}, \hat{x}_{t})$, and use the same model to output the next prediction $\hat{x}_{t+1}$. This procedure is repeated, updating the input window at each step, until the model produces $\hat{x}_{t-1+k}$. This closed-loop configuration is closer to realistic forecasting scenarios. However, because it relies on feeding back predicted values, the task typically becomes more challenging as the prediction horizon $k$ increases. This can be attributed to successive magnification of prediction errors at each step arising out of use of previously simulated predictions as training data for further prediction. Since closed-loop forecasting is implemented via iterative one-step prediction, we adopt the optimal QRC model that yields the best single-step performance. For the logistic map, our previous results show that the optimal configuration is $d=2$ and $n_{\mathrm{rep}}=2$~(see Fig.~\ref{fig3:architecture}(a)). For the H\'enon map, our previous results show that the optimal configuration is $d=1$ and $n_{\mathrm{rep}}=2$~(see Fig.~\ref{fig3:architecture}(b). Therefore, we use these settings respectively for logistic and H\'enon maps throughout, aiming to minimize the error introduced at each step and thereby maintain reliable long-horizon forecasts. \\

\begin{figure*}
    \includegraphics[width = \linewidth]{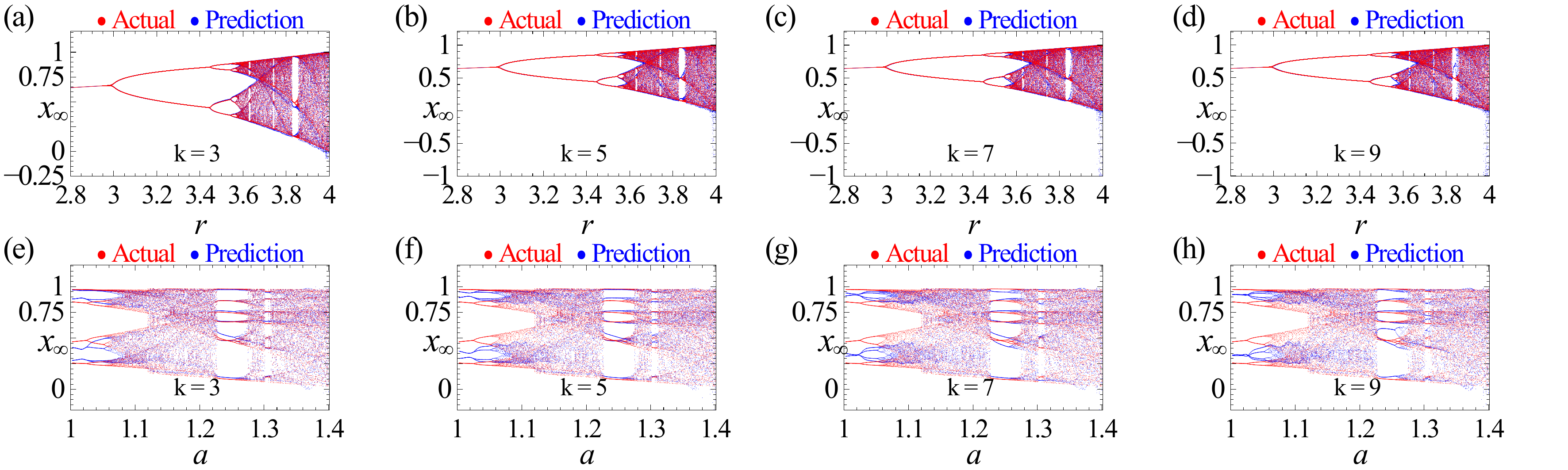}
    \caption{Actual (red) vs predicted (blue) bifurcation diagram of the logistic map  with different closed-loop prediction horizons $k$, where for (a) $k = 3$, (b) $k = 5$, (c) $k = 7$ and (d) $k = 9$. Actual (red) vs predicted (blue) bifurcation diagram of the H\'enon map with different closed-loop prediction horizons $k$, where for (e) $k = 3$, (f) $k = 5$, (g) $k = 7$, and (h) $k = 9$.}
    \label{fig4:bifurcation_with_k}
\end{figure*}

In Fig.~\ref{fig4:bifurcation_with_k}, we plot the bifurcation diagram of the logistic map (Figs.~\ref{fig4:bifurcation_with_k}(a)-(d)) and the H\'enon map (Figs.~\ref{fig4:bifurcation_with_k}(e)-(h)) obtained from multi-step forecasting using the closed-loop strategy. We display the predicted asymptotic values $x_{\infty}$ by again collecting the points from $x_{151}$ to $x_{200}$. We consider four different prediction horizons, namely $k{=}3,5,7,9$. From Fig.~\ref{fig4:bifurcation_with_k}, we observe that the model can accurately reproduce the broad bifurcation diagram of the logistic map  with respect to the control parameter $r$ (logistic map), $a$ (H\'enon map) in the fixed-point regime as well as across the periodic windows. For the logistic map, for horizon $k{=}3$, the QRC prediction yields an excellent reconstruction of the bifurcation structure even in the chaotic regime. However, when the prediction horizon increases to $k{=}5$ and beyond, the performance deteriorates in the strongly chaotic region close to $r\simeq 4$. In this regime, the iterative feedback inherent to closed-loop forecasting leads to gradual accumulation of errors, and the predicted values deviate substantially from the valid range of the logistic map, resulting in poor agreement with the true bifurcation pattern. This behavior reflects a fundamental challenge of long-horizon prediction in chaotic systems, where small inaccuracies are exponentially amplified. For the H\'enon map, the same pattern is seen, although here the disagreement between predicted values and the ground-truth is even more pronounced, which we may ultimately ascribe to the fact that the H\'enon map is a two-dimensional map with errors accumulating along both variables.    At the same time, it is worth emphasizing that our QRC employs only $d{=}2$ layers for the logistic map and only $d{=}1$ layer for the H\'enon map. Consequently, even for ${k=}3$, once the prediction is rolled forward, the model necessarily relies entirely on its own intermediate predictions (e.g., $\hat{x}_{t}$ and $\hat{x}_{t+1}$) without access to any further ground-truth values.  This highlights both the capability as well as the limitation of our QRC to perform genuine multi-step forecasting within a closed-loop setting.  \\

\begin{figure}
    \includegraphics[width = \linewidth]{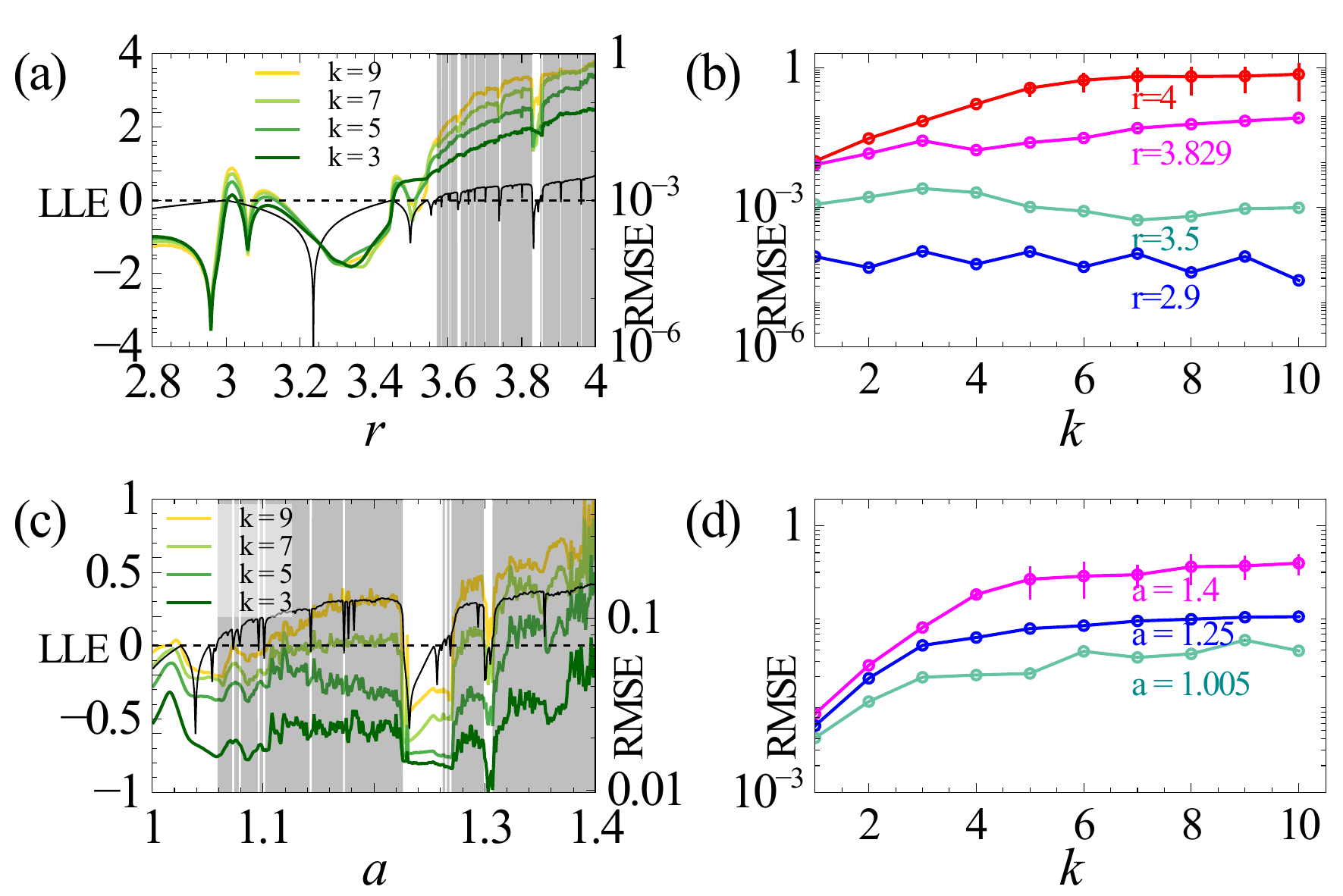}
    \caption{(a) RMSE of predicting bifurcation diagram of the logistic map with different horizon steps $k = 3,5,7,9$. LLE (black line) and chaotic regimes (grey) are depicted for reference. (b) RMSE of prediction at different regimes $r=2.9$ (fixed point, blue), $r=3.5$ (4-periodic orbit, teal), $r=3.829$ (island of stability, magenta), $r=4.0$ (chaotic, red) with prediction horizon $k$ ranging from 1 to 10. (c) same as (a) but for Henon map. (d) same as (b) but for Henon map at different regimes $a = 1.005$ (4-periodic orbit, teal) , $a = 1.25$ (7-periodic orbit, blue), $a = 1.4$ (chaotic, magenta). }
    \label{fig5:RMSE_with_k}
\end{figure}

For a quantitative understanding of the prediction error for the closed-loop strategy, we once again employ the RMSE of prediction error as a function of the map parameter $r$ (logistic map) and $a$ (H\'enon map) for various prediction horizons $k$.  In Fig.~~\ref{fig5:RMSE_with_k}(a) ((c)), we plot the largest Lyapunov exponent (LLE) together with the prediction error (RMSE) as functions of the control parameter $r$($a$) of the logistic map (H\'enon map) for different closed-loop prediction horizons $k$. We observe that before the LLE first becomes positive, i.e., before the logistic map enters the chaotic regime, the QRC achieves very similar performance across different prediction horizons $k$, all with uniformly small prediction errors. However, as $r$ increases further and the map enters the chaotic regime via successive bifurcations, the prediction error gradually grows for all prediction horizons $k$. Nevertheless, when $r$ lies within islands of stability (periodic windows embedded in chaos), the RMSE exhibits a pronounced dip compared to the surrounding chaotic region. This indicates that the QRC is highly sensitive to such stable windows and can effectively distinguish them from fully chaotic dynamics. Overall, the trend of prediction error versus logistic map parameter $r$ in the closed-loop strategy is highly consistent with the single-step prediction results reported earlier (see Fig.~\ref{fig2:prediction}). At the same time, Fig.~\ref{fig5:RMSE_with_k}(a) also indicates the prediction error increases with the forecasting horizon $k$, which can be attributed to the accumulation of prediction errors inherent in the closed-loop iterative procedure. All these results described above for the logistic map  indicating progressively worsening prediction accuracy with onset of chaos qualitatively agree with the results obtained for the H\'enon map in Fig.~\ref{fig5:RMSE_with_k}(c).\\

To further examine the closed-loop performance of the QRC across different prediction horizons, we select four representative values of the control parameter $r$, namely: (i) $r{=}2.9$, where the logistic map converges to a single fixed point; (ii) $r{=}3.5$, where the logistic map orbit exhibits periodic oscillations between four values; (iii) $r=3.829$, corresponding to an island of stability; and (iv) $r=4$, where the map is fully chaotic. For each of these cases, we generate closed-loop predictions for the time interval $x_{151}$ to $x_{200}$ under different horizons, and we repeat the test over $10$ different initial conditions. The results are reported in Fig.~\ref{fig5:RMSE_with_k}(b), where both the prediction-error and its variance (error-bars) have been depicted with respect to various prediction horizons $k$ ranging from $k{=1}$ to $k{=}10$. For the single fixed point $r=2.9$, the QRC achieves consistently accurate forecasts across all horizons, with errors $\sim 10^{-4}$. For the 4-periodic oscillation regime at $r=3.5$, where the orbit oscillates among four discrete values, the QRC likewise maintains high accuracy for all prediction horizons considered  with errors $\sim 10^{-3}$. At $r=3.829$ (island of stability), the prediction error increases moderately with the prediction horizon $k$, typically from $\sim 10^{-2}~(k{=}1)$  towards $\sim 10^{-1} ~(k{=}10)$. In contrast, for the fully chaotic case $r=4$, the forecasting error grows significantly as the horizon increases, rising from $\sim 10^{-2}~(k{=}1)$ up to $\sim 10^{0}~(k{=}6)$. For completeness, in appendix~\ref{detialofclosedloop} together with Fig.~\ref{figappendix} therein, we provide explicit comparisons between the QRC predictions and the corresponding ground-truth trajectories for prediction horizons $k{=}3,5,7,9$ at each of the four representative values of $r$ listed above. The corresponding results for the H\'enon map are depicted in Fig.~\ref{fig5:RMSE_with_k}(d) for various values of the map parameter $a$ - namely, 4-periodic orbits ($a=1.005$), 7-periodic orbits ($a=1.25$), and chaotic regime ($a=1.4$). These results again agree qualitatively with the results for the logistic map, viz., prediction error increases even more significantly with the prediction horizon $k$ within the chaotic regime. This once again highlights the intrinsic difficulty of long-horizon prediction in chaotic systems under closed-loop strategy. \\

From the results above, we conclude that using the closed-loop forecasting strategy, our QRC model can capture the distinctive properties of the logistic map across a wide range of dynamical regimes. In particular, when the logistic map is in non-chaotic regimes, the model can accurately predict the values of $x_t$ even up to relatively large prediction horizons, indicating its ability to capture the long-term behavior of the map in regimes where trajectories are stable or periodic. In the chaotic regime, the model can still reproduce the dynamics reasonably well for short prediction horizons. However, as the prediction horizon $k$ increases, accurate numerical prediction becomes extremely challenging due to the intrinsic sensitivity of chaotic dynamics and the accumulation of errors in the closed-loop strategy. Overall, our results demonstrate that the proposed QRC model possesses long-horizon closed-loop forecasting capability despite using a small input window and very low circuit depth (only two layers).  In addition, the behavior of error in long-horizon closed-loop prediction can be used for discriminating between chaotic and non-chaotic behaviors.

\section{Robustness of forecasting} So far, our results have assumed perfect unitary evolution, as well as randomized but fixed instances of reservoir Hamiltonian initialization. However, in the NISQ era, the former is a simplistic assumption as the presence of noise leads to non-unitary evolution in real setups. Moreover, the appeal of reservoir computing lies in the fact that the algorithm should work well regardless of the exact internal coupling parameters of the reservoir. Thus, for actual implementation, we must account for both these factors to benchmark the realistic predictive accuracy of our protocol. Moreover, the final expectation values readout theoretically requires infinite trials on identical ensembles to fully reconstruct corresponding probabilities, while practically we are firmly in the finite-sampling regime. Thus, firstly, we show that by training the system with the measured data from a noisy reservoir we can reach significant robustness against dephasing. Secondly, we demonstrate with statistical analysis that our protocol is robust against randomness in reservoir couplings. Finally, we show that the a relatively small number of measurements for readout already suffices to agree with theoretical results reported here.\\

\subsection{Robustness against decoherence}  
\begin{figure}
    \includegraphics[width = \linewidth]{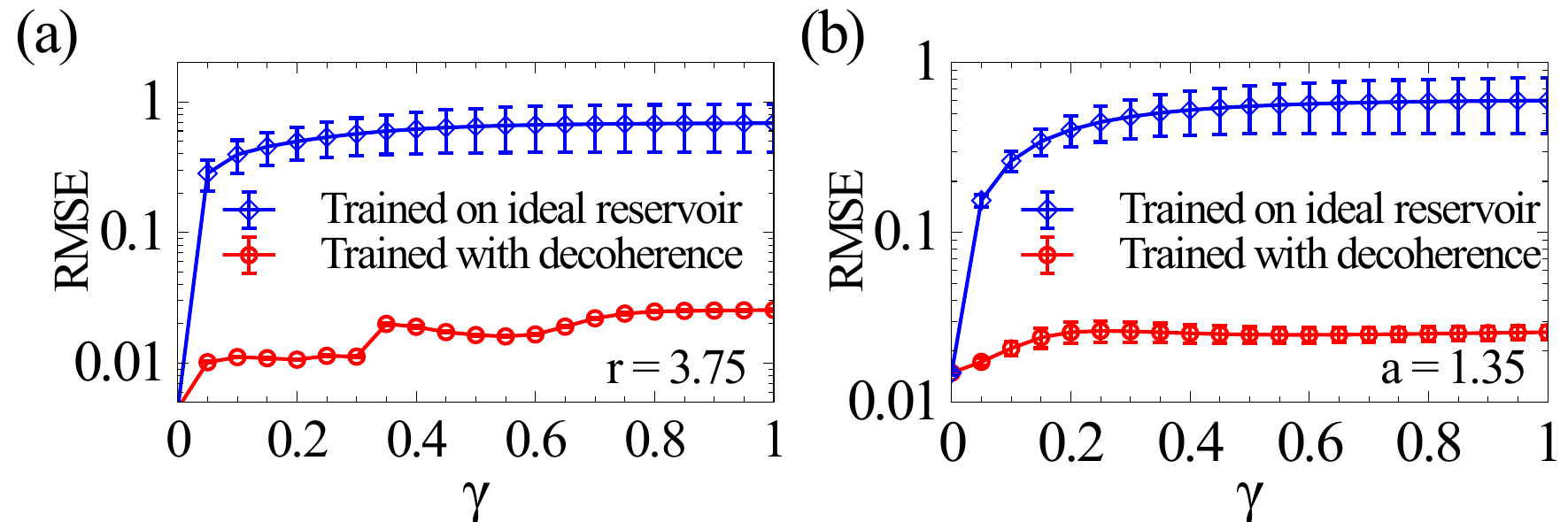}
    \caption{The RMSE of QRC with decoherent noise strength $\gamma$ for (a) the logistic map, and (b) the H\'enon map. Blue (red) lines indicate training without (with) decoherence.  }
    \label{fig6:decoherence}
\end{figure}

So far, we assumed a perfect unitary evolution of the reservoir.  However, in practice, unwanted interaction with the environment is unavoidable. Such interaction generally leads to decoherence, i.e., gradual loss of quantum properties in a physical system after some time, which is the biggest challenge to physical implementation of quantum algorithms \cite{nielsen2010quantum}. Dephasing is the most common decoherence effect whose origin is the presence of random site-dependent magnetic fields in the environment. The effect of such fields is to change the unitary evolution into a Lindblad master equation described by   
\begin{equation}
    \frac{\mathrm{d}\rho}{\mathrm{d}\tau}= -i~[H,\rho] + \gamma \sum_{k}\left( Z_k\rho Z_k^\dagger-\rho \right).
\end{equation}
where $\gamma$ represents the strength of the dephasing~\cite{nielsen2010quantum}.  Here, we evaluate the performance of the QRC framework under phase damping noise using two different approaches. In the first approach, we train the QRC in an ideal (noise-free) environment but test it in the presence of noise. However, in the second approach, training and testing the QRC are both performed under the same noise. As shown in Figs.~\ref{fig6:decoherence}(a) and (b), the predictive accuracy of the first model rapidly declines with noise strength $\gamma$ for both the logistic and the H\'enon maps, respectively. This decline comes as no surprise as our measurement readout is along on the Pauli-X basis, thus the information retrieved from the measurements also decays exponentially as the coherence decreases. However, surprisingly, the second approach performs exceptionally well and maintains robustness in accurate prediction even under reasonably large noise, leading to orders of magnitude better performance for both logistic and H\'enon maps. These results strongly indicate in situ training is preferable in the QRC framework. Physically, this stems from the fact the predictor trained in situ succeeds in learning both the noise parameter $\gamma$ as well as the time-series data encoded, while the predictor trained on a noise-free environment naturally fails to account for the noise parameter $\gamma$ as a background parameter for the test data. \\

\begin{figure}
    \includegraphics[width = \linewidth]{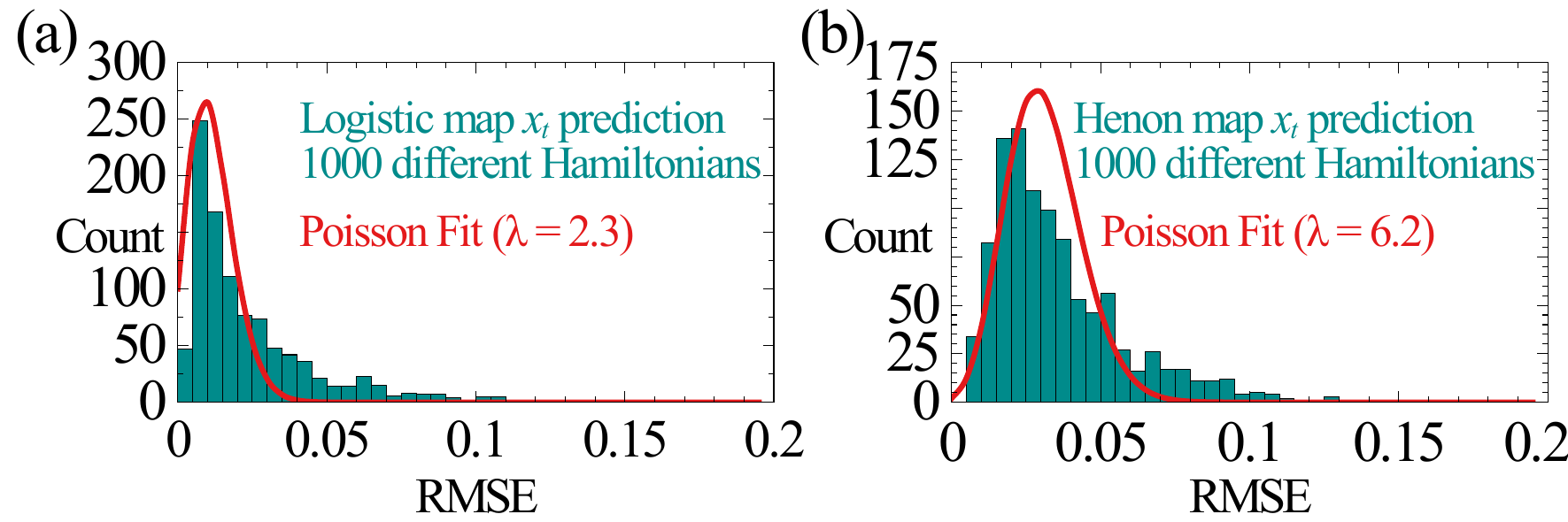}
    \caption{Histogram for RMSE statistics of 1000 randomly chosen Hamiltonian reservoirs for (a) logistic map, and (b) H\'enon map. Red lines are Poisson distribution fits for $n_{\textrm{bin}}{=}40$ bins.}
    \label{fig7:histogram}
\end{figure}

\begin{figure}
    \includegraphics[width = \linewidth]{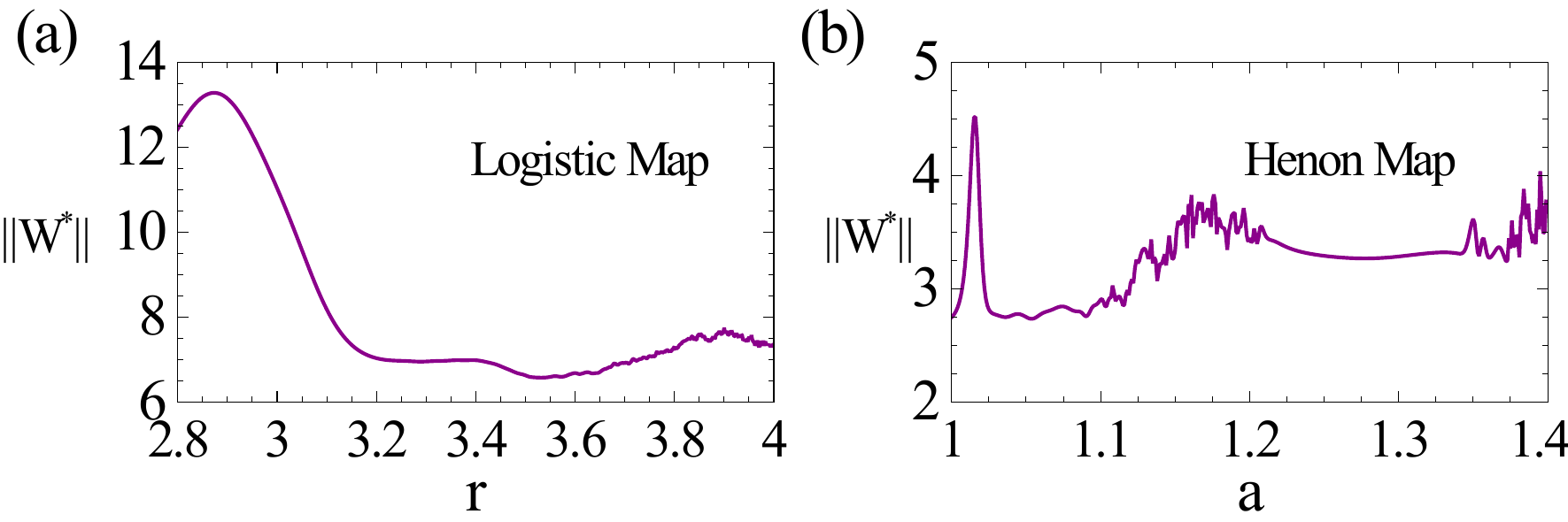}
    \caption{Norm $||W^*||$ of optimal weight matrix $W^*$ of QRC for the logistic map (left), and the H\'enon map (right).  }
    \label{fig8:normW}
\end{figure}

\subsection{Robustness against reservoir randomness} 
\label{engineering}
In this section, we discuss the robustness of our QRC protocol for predicting chaotic maps with respect to the choice of the parameters in the quantum reservoir. In general, the lack of sensitivity to such parameterization is a highly desirable trait for reservoir computing, where the explicit goal is to only adjust the weight matrix $\textbf{W}$ at the end. To check this, we randomly selected $1000$ Hamiltonians for both the logistic and the H\'enon maps and binning their respective predictive errors, represented by RMSEs, in the histograms of Fig.~\ref{fig7:histogram}. The results indicate that the majority of Hamiltonians yield low RMSE, with only a small fraction exhibiting big errors. Moreover, in both cases, the histograms are Poissonian in nature, which means only a very small fraction of all the Hamiltonians are `problematic' with the vast majority of randomly chosen Hamiltonians yielding close to maximum predictive accuracy. 
This is consistent with the measure concentration phenomenon encountered in locally entangled many-body systems such as the present reservoir based on transverse XY spin chains, where most dynamical states are in fact concentrated within a small subspace of the overall Hilbert space ~\cite{dubey2012approach, ithier2017dynamical, reimann2018dynamical}. While an explicit quantification is beyond the scope of the present work, it appears randomly sampled Hamiltonians do typically arise from this subspace and the exceptions outside this subspace are rare, thus a Poissonian treatment is apt.

\subsection{Robustness against finite sampling error}
In the numerical simulations presented above, we directly computed the expectation values of Pauli-$X$ observables at the last layer. However, these expectation values are theoretically only valid in the asymptotic limit of large number of trials. In an experimental setting, however, these expectation values must be estimated from a finite number of measurement shots. Assuming that each observable $X$ is measured with $M$ shots and independent measurements on identically prepared samples, the estimated expectation value $\langle X \rangle_\mathrm{sam}$ is centered around the true expectation value $\langle X \rangle$  with an width $\Delta(X)$, i.e., 
\begin{equation}
    \langle X \rangle = \langle X \rangle_\mathrm{sam} \pm \Delta(X),
\end{equation}
where the statistical estimation error $ \Delta(X) {\sim}\sqrt{\frac{\langle X^2\rangle-\langle X \rangle^2}{M}}$ diminishes as ${\sim} \mathcal{O}(1/\sqrt{M})$. We note that since the Pauli-$X$ operator satisfies $\langle X^2\rangle=1$ and $0\leq\langle X \rangle^2\leq 1$ throughout, the numerator $\langle X^2\rangle - \langle X\rangle^2$ is bounded everywhere. \\

We now note that the prediction from the QRC is obtained via a linear readout of the quantum measured features, i.e., 
\begin{equation}
\hat{y} = W\vec{m},
\end{equation}
where $\vec{m}{=}[\langle X_1\rangle, \langle X_2\rangle, \cdots, \langle X_{n_\textrm{I}+n_\textrm{H}}\rangle]^T$. Writing $\vec{m}=\vec{m}_{\mathrm{sam}}\pm\vec{m_{\Delta}}$, where $\vec{m_{\Delta}}{=}[\Delta( X_1), \Delta(X_2), \cdots,\Delta (X_{n_\textrm{I}+n_\textrm{H}})]^T$ yields
\begin{equation}
\hat{y}=W(\vec{m}_{\mathrm{sam}}\pm\vec{m_{\Delta}}).
\end{equation} \\
In other words, the finite-sampling error in the measured feature vector $\vec{m}$ propagates linearly into the prediction. Practically, we want the prediction deviation induced by this finite sampling error to be below some tolerance $\epsilon$, namely:
\begin{equation}
\|W\vec{m}_{\mathrm{sam}}-W\vec{m}\|_2=\|W\vec{m_{\Delta}}\|_2<\epsilon.
\end{equation}
Now, vide the Cauchy-Schwarz inequality, we already know that
\begin{equation}
\|W\vec{m_{\Delta}}\|_2\leq \|W\|_2\|\vec{m_{\Delta}}\|_2<\epsilon.
\end{equation}
Because $\|\vec{m_{\Delta}}\|_2=\sqrt{\sum_i^{n_\textrm{I}+n_\textrm{H}}\frac{1-\langle X_i\rangle}{M}}\leq\sqrt{\frac{n_\textrm{I}+n_\textrm{H}}{M}}$, this inequality yields an explicit relationship between the required shot number $M$ and the desired tolerance $\epsilon$ as $\frac{(n_\textrm{I}+n_\textrm{H})\|W\|_2}{\epsilon^2}\leq M$. In particular, this analysis indicates that it is advantageous for the readout weights to have a moderate norm; \emph{i.e.}, the norm $\|W^*\|^2$ corresponding to the optimal weight matrix $W^*$ arrived at from the training should not be too large in order to avoid excessive amplification of finite-sampling error. In Fig.~\ref{fig8:normW}, we plot $\|W^*\|^2$ as a function of map parameters ($r$ for the logistic map and $a$ for the H\'enon map). In both cases, We find that $\|W^*\|^2$ remains modest across the parameter range, suggesting that an experimentally feasible number of measurement shots should be sufficient to overcome the finite sampling error. \\

Thus, we have demonstrated the robustness of QRC against coherent and incoherent noise, as well as finite sampling errors. At the same time, we note that several recent works on QRC have aimed to further enhance noise resilience by incorporating error-mitigation-inspired strategies, for example, by modifying the training protocol~\cite{ahmedOptimalTrainingFinitely2025} and/or the measurement scheme~\cite{mujalTimeseriesQuantumReservoir2023}. This line of research is complementary to the results presented here, as such techniques may further improve the predictive accuracy and stability of QRC, especially for long-horizon forecasting tasks.

\section{Conclusion} 
In this work, we have proposed a QRC algorithm for predicting and characterizing nonlinear dynamics in discrete  maps with chaotic regimes. We have demonstrated the effectiveness of our method on the logistic and H\'enon maps. While we can achieve accurate prediction about the future steps of our time series, we have also been able to optimize the QRC encoding protocol based on the features of nonlinear maps studied herein, paving the way for further optimization of the QRC architecure for experimental time series data.  
In addition, through closed-loop prediction of distant future steps one can discriminate non-chaotic versus chaotic phases. In this situation, while the closed-loop strategy can faithfully predict distant future outcomes in non-chaotic phases, in the chaotic phase the error grows significantly as the horizon for future steps increases.
The proposed QRC algorithm is resilient against decoherence and reservoir randomness, making it suitable for near-term quantum devices, and relies on reasonably small number of measurement samples. We note here that in contrast to other recent proposals of applying QRC to chaotic dynamics forecasting \cite{wang2024enhanced,ahmed2025robust} studying continuous dynamical evolution, we study discrete chaotic maps. Moreover, we use simple phase encoding as opposed to computationally costly amplitude encoding \cite{wang2024enhanced}; and employ a simple reservoir only requiring nearest neighbor XX interaction which is readily available in superconducting quantum simulators~\cite{Xu_2018,Huang_2020,Blais_2021,Gong2021Experimental}, ion-trap systems~\cite{blatt2012quantum,monroe2021programmable,katz2023programmable} and optical lattices~\cite{noh2016quantum,wang2019simulating,supranoExperimentalPropertyReconstruction2024} as opposed to requiring complex reservoir engineering with fully-connected~\cite{ahmed2025robust} or non-sparse Hamiltonians~\cite{wang2024enhanced}.  \\

\section{Acknowledgments}  
AB acknowledges support from the National Natural Science Foundation of China (grants No. 12274059, No. 12574528, No. 1251101297 and No. W2541020). LM gratefully acknowledges the support of the ``Hundred Talents'' program of the University of Electronic Science and Technology of China, the ``Outstanding Young Talents Program (Overseas)'' of the National Natural Science Foundation of China, and the talent programs of the Sichuan province and Chengdu municipality. \\

\appendix

\section{Closed-loop prediction with different prediction horizons for logistic and H\'enon maps}
\label{detialofclosedloop}
\begin{figure*}
    \includegraphics[width = 0.8\linewidth]{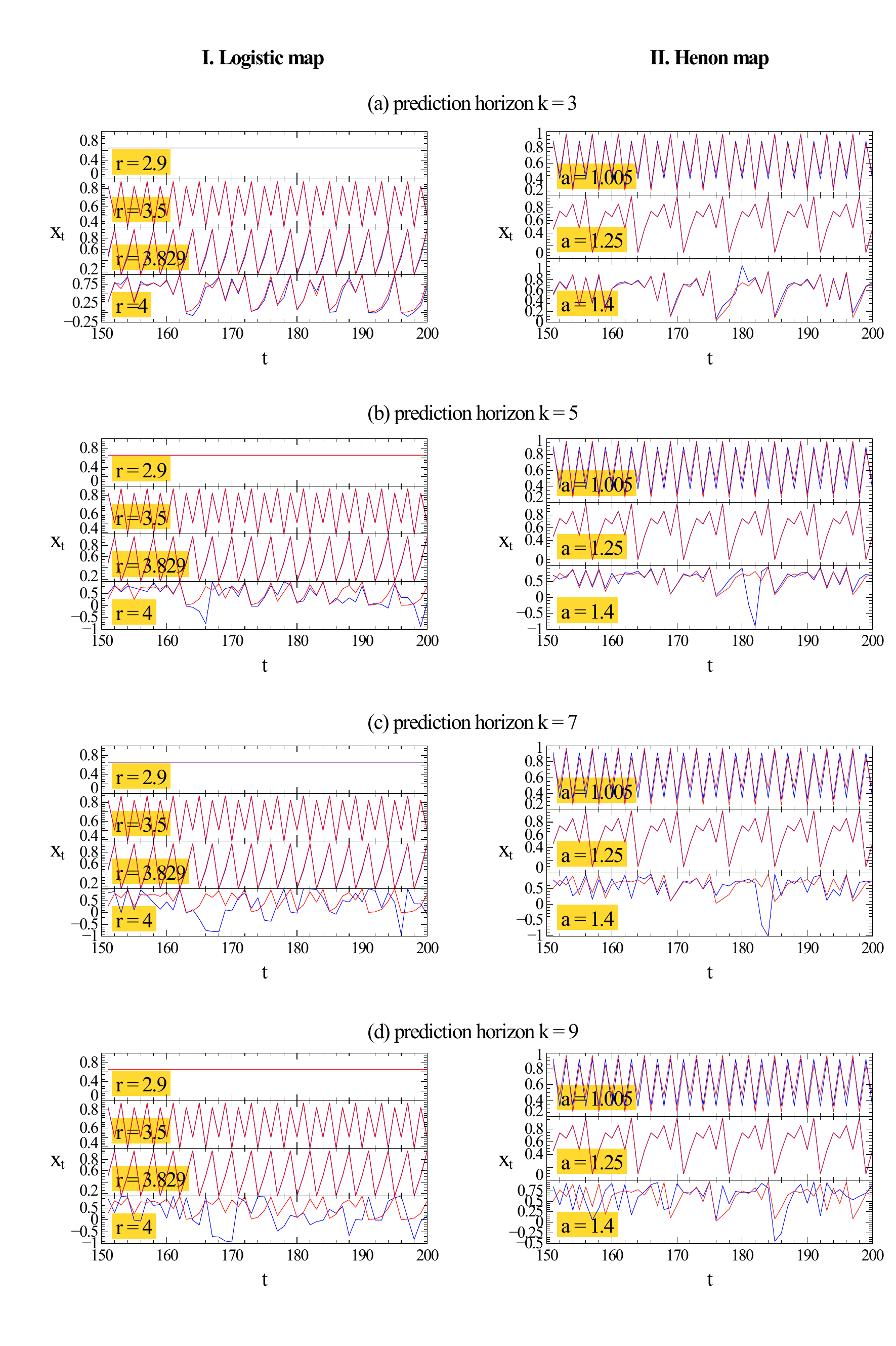}
    \caption{Closed-loop prediction with different horizons $k$ for logistic map (left) and Henon map (right) for time-steps $t\in[151,200]$. Blue lines represent predicted values and red lines represent actual values. For logistic map $r = 2.9$ (fixed point), $r = 3.5,3.829$ (periodic orbits) and $r=4$ (chaotic) map parameters have been considered. For H\'enon map, $a=1.005,1.25$(periodic orbits) and $a=1.4$ (chaotic) map parameters have been considered. }
    \label{figappendix}
\end{figure*}

In this Appendix, we report the prediction results for the logistic and H\'enon maps under different closed-loop horizons (see Fig.~\ref{figappendix}). For the logistic map, we focus on four representative values of the control parameter $r$ ($r=2.9, 3.5, 3.829, 4$ as specified in the main text).  For the H\'enon map, we focus on three representative values of the control parameter $a$ ($a=1.005,1.25,1.4$ as specified in the main text).  For the non-chaotic cases of both maps, the QRC model faithfully reproduces the corresponding trajectories, indicating a reliable long-horizon rollout in stable and periodic regimes. For the chaotic regimes of both maps, we find that the model can still provide a reasonably good forecast for a short horizon, in particular for $k=3$. However, as the horizon is further increased, the predictions become progressively less reliable due to the accumulation of errors and the intrinsic sensitivity of chaotic dynamics. \\

\textbf{Code Availability.}  The Code used in this paper can be found  \href{https://github.com/LeeQY1996/Quantum-Reservoir-Computing-for-Predicting-Chaotic-Maps}{here}  in the Github Repositories. \\

\bibliography{qrl_chaotic}

%merlin.mbs apsrev4-1.bst 2010-07-25 4.21a (PWD, AO, DPC) hacked
%Control: key (0)
%Control: author (72) initials jnrlst
%Control: editor formatted (1) identically to author
%Control: production of article title (-1) disabled
%Control: page (0) single
%Control: year (1) truncated
%Control: production of eprint (0) enabled
\begin{thebibliography}{73}%
\makeatletter
\providecommand \@ifxundefined [1]{%
 \@ifx{#1\undefined}
}%
\providecommand \@ifnum [1]{%
 \ifnum #1\expandafter \@firstoftwo
 \else \expandafter \@secondoftwo
 \fi
}%
\providecommand \@ifx [1]{%
 \ifx #1\expandafter \@firstoftwo
 \else \expandafter \@secondoftwo
 \fi
}%
\providecommand \natexlab [1]{#1}%
\providecommand \enquote  [1]{``#1''}%
\providecommand \bibnamefont  [1]{#1}%
\providecommand \bibfnamefont [1]{#1}%
\providecommand \citenamefont [1]{#1}%
\providecommand \href@noop [0]{\@secondoftwo}%
\providecommand \href [0]{\begingroup \@sanitize@url \@href}%
\providecommand \@href[1]{\@@startlink{#1}\@@href}%
\providecommand \@@href[1]{\endgroup#1\@@endlink}%
\providecommand \@sanitize@url [0]{\catcode `\\12\catcode `\$12\catcode `\&12\catcode `\#12\catcode `\^12\catcode `\_12\catcode `\%12\relax}%
\providecommand \@@startlink[1]{}%
\providecommand \@@endlink[0]{}%
\providecommand \url  [0]{\begingroup\@sanitize@url \@url }%
\providecommand \@url [1]{\endgroup\@href {#1}{\urlprefix }}%
\providecommand \urlprefix  [0]{URL }%
\providecommand \Eprint [0]{\href }%
\providecommand \doibase [0]{http://dx.doi.org/}%
\providecommand \selectlanguage [0]{\@gobble}%
\providecommand \bibinfo  [0]{\@secondoftwo}%
\providecommand \bibfield  [0]{\@secondoftwo}%
\providecommand \translation [1]{[#1]}%
\providecommand \BibitemOpen [0]{}%
\providecommand \bibitemStop [0]{}%
\providecommand \bibitemNoStop [0]{.\EOS\space}%
\providecommand \EOS [0]{\spacefactor3000\relax}%
\providecommand \BibitemShut  [1]{\csname bibitem#1\endcsname}%
\let\auto@bib@innerbib\@empty
%</preamble>
\bibitem [{\citenamefont {Shor}(1994)}]{shorAlgorithmsQuantumComputation1994}%
  \BibitemOpen
  \bibfield  {author} {\bibinfo {author} {\bibfnamefont {P.}~\bibnamefont {Shor}},\ }in\ \href {\doibase 10.1109/SFCS.1994.365700} {\emph {\bibinfo {booktitle} {Proceedings 35th {{Annual Symposium}} on {{Foundations}} of {{Computer Science}}}}}\ (\bibinfo {year} {1994})\ pp.\ \bibinfo {pages} {124--134}\BibitemShut {NoStop}%
\bibitem [{\citenamefont {Vandersypen}\ \emph {et~al.}(2001)\citenamefont {Vandersypen}, \citenamefont {Steffen}, \citenamefont {Breyta}, \citenamefont {Yannoni}, \citenamefont {Sherwood},\ and\ \citenamefont {Chuang}}]{vandersypenExperimentalRealizationShors2001}%
  \BibitemOpen
  \bibfield  {author} {\bibinfo {author} {\bibfnamefont {L.~M.~K.}\ \bibnamefont {Vandersypen}}, \bibinfo {author} {\bibfnamefont {M.}~\bibnamefont {Steffen}}, \bibinfo {author} {\bibfnamefont {G.}~\bibnamefont {Breyta}}, \bibinfo {author} {\bibfnamefont {C.~S.}\ \bibnamefont {Yannoni}}, \bibinfo {author} {\bibfnamefont {M.~H.}\ \bibnamefont {Sherwood}}, \ and\ \bibinfo {author} {\bibfnamefont {I.~L.}\ \bibnamefont {Chuang}},\ }\href {\doibase 10.1038/414883a} {\bibfield  {journal} {\bibinfo  {journal} {Nature}\ }\textbf {\bibinfo {volume} {414}},\ \bibinfo {pages} {883} (\bibinfo {year} {2001})}\BibitemShut {NoStop}%
\bibitem [{\citenamefont {Grover}(1996)}]{groverFastQuantumMechanical1996}%
  \BibitemOpen
  \bibfield  {author} {\bibinfo {author} {\bibfnamefont {L.~K.}\ \bibnamefont {Grover}},\ }in\ \href {\doibase 10.1145/237814.237866} {\emph {\bibinfo {booktitle} {Proceedings of the Twenty-Eighth Annual {{ACM}} Symposium on {{Theory}} of Computing - {{STOC}} '96}}}\ (\bibinfo  {publisher} {ACM Press},\ \bibinfo {address} {Philadelphia, Pennsylvania, United States},\ \bibinfo {year} {1996})\ pp.\ \bibinfo {pages} {212--219}\BibitemShut {NoStop}%
\bibitem [{\citenamefont {Grover}(1998)}]{groverQuantumComputersCan1998}%
  \BibitemOpen
  \bibfield  {author} {\bibinfo {author} {\bibfnamefont {L.~K.}\ \bibnamefont {Grover}},\ }\href {\doibase 10.1103/PhysRevLett.80.4329} {\bibfield  {journal} {\bibinfo  {journal} {Physical Review Letters}\ }\textbf {\bibinfo {volume} {80}},\ \bibinfo {pages} {4329} (\bibinfo {year} {1998})}\BibitemShut {NoStop}%
\bibitem [{\citenamefont {Preskill}(2018)}]{preskillQuantumComputingNISQ2018}%
  \BibitemOpen
  \bibfield  {author} {\bibinfo {author} {\bibfnamefont {J.}~\bibnamefont {Preskill}},\ }\href {\doibase 10.22331/q-2018-08-06-79} {\bibfield  {journal} {\bibinfo  {journal} {Quantum}\ }\textbf {\bibinfo {volume} {2}},\ \bibinfo {pages} {79} (\bibinfo {year} {2018})},\ \Eprint {http://arxiv.org/abs/1801.00862} {arXiv:1801.00862 [cond-mat, physics:quant-ph]} \BibitemShut {NoStop}%
\bibitem [{\citenamefont {Banchi}(2024)}]{Banchi_2024}%
  \BibitemOpen
  \bibfield  {author} {\bibinfo {author} {\bibfnamefont {L.}~\bibnamefont {Banchi}},\ }\href {\doibase 10.1088/2632-2153/ad444a} {\bibfield  {journal} {\bibinfo  {journal} {Machine Learning: Science and Technology}\ }\textbf {\bibinfo {volume} {5}},\ \bibinfo {pages} {025036} (\bibinfo {year} {2024})}\BibitemShut {NoStop}%
\bibitem [{\citenamefont {Khosrojerdi}\ \emph {et~al.}(2025)\citenamefont {Khosrojerdi}, \citenamefont {Pereira}, \citenamefont {Cuccoli},\ and\ \citenamefont {Banchi}}]{Khosrojerdi_2025}%
  \BibitemOpen
  \bibfield  {author} {\bibinfo {author} {\bibfnamefont {M.}~\bibnamefont {Khosrojerdi}}, \bibinfo {author} {\bibfnamefont {J.~L.}\ \bibnamefont {Pereira}}, \bibinfo {author} {\bibfnamefont {A.}~\bibnamefont {Cuccoli}}, \ and\ \bibinfo {author} {\bibfnamefont {L.}~\bibnamefont {Banchi}},\ }\href {\doibase 10.1088/2058-9565/ada79b} {\bibfield  {journal} {\bibinfo  {journal} {Quantum Science and Technology}\ }\textbf {\bibinfo {volume} {10}},\ \bibinfo {pages} {025006} (\bibinfo {year} {2025})}\BibitemShut {NoStop}%
\bibitem [{\citenamefont {Zimbor{\'a}s}\ \emph {et~al.}(2025)\citenamefont {Zimbor{\'a}s}, \citenamefont {Koczor}, \citenamefont {Holmes}, \citenamefont {Borrelli}, \citenamefont {Gily{\'e}n}, \citenamefont {Huang}, \citenamefont {Cai}, \citenamefont {Ac{\'\i}n}, \citenamefont {Aolita}, \citenamefont {Banchi} \emph {et~al.}}]{zimboras2025myths}%
  \BibitemOpen
  \bibfield  {author} {\bibinfo {author} {\bibfnamefont {Z.}~\bibnamefont {Zimbor{\'a}s}}, \bibinfo {author} {\bibfnamefont {B.}~\bibnamefont {Koczor}}, \bibinfo {author} {\bibfnamefont {Z.}~\bibnamefont {Holmes}}, \bibinfo {author} {\bibfnamefont {E.-M.}\ \bibnamefont {Borrelli}}, \bibinfo {author} {\bibfnamefont {A.}~\bibnamefont {Gily{\'e}n}}, \bibinfo {author} {\bibfnamefont {H.-Y.}\ \bibnamefont {Huang}}, \bibinfo {author} {\bibfnamefont {Z.}~\bibnamefont {Cai}}, \bibinfo {author} {\bibfnamefont {A.}~\bibnamefont {Ac{\'\i}n}}, \bibinfo {author} {\bibfnamefont {L.}~\bibnamefont {Aolita}}, \bibinfo {author} {\bibfnamefont {L.}~\bibnamefont {Banchi}},  \emph {et~al.},\ }\href@noop {} {\bibfield  {journal} {\bibinfo  {journal} {arXiv preprint arXiv:2501.05694}\ } (\bibinfo {year} {2025})}\BibitemShut {NoStop}%
\bibitem [{\citenamefont {Peruzzo}\ \emph {et~al.}(2014)\citenamefont {Peruzzo}, \citenamefont {McClean}, \citenamefont {Shadbolt}, \citenamefont {Yung}, \citenamefont {Zhou}, \citenamefont {Love}, \citenamefont {Aspuru-Guzik},\ and\ \citenamefont {O’Brien}}]{Peruzzo_2014}%
  \BibitemOpen
  \bibfield  {author} {\bibinfo {author} {\bibfnamefont {A.}~\bibnamefont {Peruzzo}}, \bibinfo {author} {\bibfnamefont {J.}~\bibnamefont {McClean}}, \bibinfo {author} {\bibfnamefont {P.}~\bibnamefont {Shadbolt}}, \bibinfo {author} {\bibfnamefont {M.-H.}\ \bibnamefont {Yung}}, \bibinfo {author} {\bibfnamefont {X.-Q.}\ \bibnamefont {Zhou}}, \bibinfo {author} {\bibfnamefont {P.~J.}\ \bibnamefont {Love}}, \bibinfo {author} {\bibfnamefont {A.}~\bibnamefont {Aspuru-Guzik}}, \ and\ \bibinfo {author} {\bibfnamefont {J.~L.}\ \bibnamefont {O’Brien}},\ }\href {\doibase 10.1038/ncomms5213} {\bibfield  {journal} {\bibinfo  {journal} {Nature Communications}\ }\textbf {\bibinfo {volume} {5}} (\bibinfo {year} {2014}),\ 10.1038/ncomms5213}\BibitemShut {NoStop}%
\bibitem [{\citenamefont {Bharti}\ \emph {et~al.}(2022)\citenamefont {Bharti}, \citenamefont {{Cervera-Lierta}}, \citenamefont {Kyaw}, \citenamefont {Haug}, \citenamefont {{Alperin-Lea}}, \citenamefont {Anand}, \citenamefont {Degroote}, \citenamefont {Heimonen}, \citenamefont {Kottmann}, \citenamefont {Menke}, \citenamefont {Mok}, \citenamefont {Sim}, \citenamefont {Kwek},\ and\ \citenamefont {{Aspuru-Guzik}}}]{bhartiNoisyIntermediatescaleQuantum2022}%
  \BibitemOpen
  \bibfield  {author} {\bibinfo {author} {\bibfnamefont {K.}~\bibnamefont {Bharti}}, \bibinfo {author} {\bibfnamefont {A.}~\bibnamefont {{Cervera-Lierta}}}, \bibinfo {author} {\bibfnamefont {T.~H.}\ \bibnamefont {Kyaw}}, \bibinfo {author} {\bibfnamefont {T.}~\bibnamefont {Haug}}, \bibinfo {author} {\bibfnamefont {S.}~\bibnamefont {{Alperin-Lea}}}, \bibinfo {author} {\bibfnamefont {A.}~\bibnamefont {Anand}}, \bibinfo {author} {\bibfnamefont {M.}~\bibnamefont {Degroote}}, \bibinfo {author} {\bibfnamefont {H.}~\bibnamefont {Heimonen}}, \bibinfo {author} {\bibfnamefont {J.~S.}\ \bibnamefont {Kottmann}}, \bibinfo {author} {\bibfnamefont {T.}~\bibnamefont {Menke}}, \bibinfo {author} {\bibfnamefont {W.-K.}\ \bibnamefont {Mok}}, \bibinfo {author} {\bibfnamefont {S.}~\bibnamefont {Sim}}, \bibinfo {author} {\bibfnamefont {L.-C.}\ \bibnamefont {Kwek}}, \ and\ \bibinfo {author} {\bibfnamefont {A.}~\bibnamefont {{Aspuru-Guzik}}},\ }\href {\doibase 10.1103/RevModPhys.94.015004} {\bibfield  {journal} {\bibinfo  {journal}
  {Reviews of Modern Physics}\ }\textbf {\bibinfo {volume} {94}},\ \bibinfo {pages} {015004} (\bibinfo {year} {2022})}\BibitemShut {NoStop}%
\bibitem [{\citenamefont {Yuan}\ \emph {et~al.}(2019)\citenamefont {Yuan}, \citenamefont {Endo}, \citenamefont {Zhao}, \citenamefont {Li},\ and\ \citenamefont {Benjamin}}]{Yuan_2019}%
  \BibitemOpen
  \bibfield  {author} {\bibinfo {author} {\bibfnamefont {X.}~\bibnamefont {Yuan}}, \bibinfo {author} {\bibfnamefont {S.}~\bibnamefont {Endo}}, \bibinfo {author} {\bibfnamefont {Q.}~\bibnamefont {Zhao}}, \bibinfo {author} {\bibfnamefont {Y.}~\bibnamefont {Li}}, \ and\ \bibinfo {author} {\bibfnamefont {S.~C.}\ \bibnamefont {Benjamin}},\ }\href {\doibase 10.22331/q-2019-10-07-191} {\bibfield  {journal} {\bibinfo  {journal} {Quantum}\ }\textbf {\bibinfo {volume} {3}},\ \bibinfo {pages} {191} (\bibinfo {year} {2019})}\BibitemShut {NoStop}%
\bibitem [{\citenamefont {Tilly}\ \emph {et~al.}(2022)\citenamefont {Tilly}, \citenamefont {Chen}, \citenamefont {Cao}, \citenamefont {Picozzi}, \citenamefont {Setia}, \citenamefont {Li}, \citenamefont {Grant}, \citenamefont {Wossnig}, \citenamefont {Rungger}, \citenamefont {Booth},\ and\ \citenamefont {Tennyson}}]{Tilly_2022}%
  \BibitemOpen
  \bibfield  {author} {\bibinfo {author} {\bibfnamefont {J.}~\bibnamefont {Tilly}}, \bibinfo {author} {\bibfnamefont {H.}~\bibnamefont {Chen}}, \bibinfo {author} {\bibfnamefont {S.}~\bibnamefont {Cao}}, \bibinfo {author} {\bibfnamefont {D.}~\bibnamefont {Picozzi}}, \bibinfo {author} {\bibfnamefont {K.}~\bibnamefont {Setia}}, \bibinfo {author} {\bibfnamefont {Y.}~\bibnamefont {Li}}, \bibinfo {author} {\bibfnamefont {E.}~\bibnamefont {Grant}}, \bibinfo {author} {\bibfnamefont {L.}~\bibnamefont {Wossnig}}, \bibinfo {author} {\bibfnamefont {I.}~\bibnamefont {Rungger}}, \bibinfo {author} {\bibfnamefont {G.~H.}\ \bibnamefont {Booth}}, \ and\ \bibinfo {author} {\bibfnamefont {J.}~\bibnamefont {Tennyson}},\ }\href {\doibase 10.1016/j.physrep.2022.08.003} {\bibfield  {journal} {\bibinfo  {journal} {Physics Reports}\ }\textbf {\bibinfo {volume} {986}},\ \bibinfo {pages} {1–128} (\bibinfo {year} {2022})}\BibitemShut {NoStop}%
\bibitem [{\citenamefont {Cerezo}\ \emph {et~al.}(2021)\citenamefont {Cerezo}, \citenamefont {Arrasmith}, \citenamefont {Babbush}, \citenamefont {Benjamin}, \citenamefont {Endo}, \citenamefont {Fujii}, \citenamefont {McClean}, \citenamefont {Mitarai}, \citenamefont {Yuan}, \citenamefont {Cincio},\ and\ \citenamefont {Coles}}]{cerezoVariationalQuantumAlgorithms2021}%
  \BibitemOpen
  \bibfield  {author} {\bibinfo {author} {\bibfnamefont {M.}~\bibnamefont {Cerezo}}, \bibinfo {author} {\bibfnamefont {A.}~\bibnamefont {Arrasmith}}, \bibinfo {author} {\bibfnamefont {R.}~\bibnamefont {Babbush}}, \bibinfo {author} {\bibfnamefont {S.~C.}\ \bibnamefont {Benjamin}}, \bibinfo {author} {\bibfnamefont {S.}~\bibnamefont {Endo}}, \bibinfo {author} {\bibfnamefont {K.}~\bibnamefont {Fujii}}, \bibinfo {author} {\bibfnamefont {J.~R.}\ \bibnamefont {McClean}}, \bibinfo {author} {\bibfnamefont {K.}~\bibnamefont {Mitarai}}, \bibinfo {author} {\bibfnamefont {X.}~\bibnamefont {Yuan}}, \bibinfo {author} {\bibfnamefont {L.}~\bibnamefont {Cincio}}, \ and\ \bibinfo {author} {\bibfnamefont {P.~J.}\ \bibnamefont {Coles}},\ }\href {\doibase 10.1038/s42254-021-00348-9} {\bibfield  {journal} {\bibinfo  {journal} {Nature Reviews Physics}\ }\textbf {\bibinfo {volume} {3}},\ \bibinfo {pages} {625} (\bibinfo {year} {2021})}\BibitemShut {NoStop}%
\bibitem [{\citenamefont {Li}\ \emph {et~al.}(2023)\citenamefont {Li}, \citenamefont {Mukhopadhyay},\ and\ \citenamefont {Bayat}}]{Li_2023}%
  \BibitemOpen
  \bibfield  {author} {\bibinfo {author} {\bibfnamefont {Q.}~\bibnamefont {Li}}, \bibinfo {author} {\bibfnamefont {C.}~\bibnamefont {Mukhopadhyay}}, \ and\ \bibinfo {author} {\bibfnamefont {A.}~\bibnamefont {Bayat}},\ }\href {\doibase 10.1103/physrevresearch.5.043175} {\bibfield  {journal} {\bibinfo  {journal} {Physical Review Research}\ }\textbf {\bibinfo {volume} {5}} (\bibinfo {year} {2023}),\ 10.1103/physrevresearch.5.043175}\BibitemShut {NoStop}%
\bibitem [{\citenamefont {McClean}\ \emph {et~al.}(2018)\citenamefont {McClean}, \citenamefont {Boixo}, \citenamefont {Smelyanskiy}, \citenamefont {Babbush},\ and\ \citenamefont {Neven}}]{mccleanBarrenPlateausQuantum2018}%
  \BibitemOpen
  \bibfield  {author} {\bibinfo {author} {\bibfnamefont {J.~R.}\ \bibnamefont {McClean}}, \bibinfo {author} {\bibfnamefont {S.}~\bibnamefont {Boixo}}, \bibinfo {author} {\bibfnamefont {V.~N.}\ \bibnamefont {Smelyanskiy}}, \bibinfo {author} {\bibfnamefont {R.}~\bibnamefont {Babbush}}, \ and\ \bibinfo {author} {\bibfnamefont {H.}~\bibnamefont {Neven}},\ }\href {\doibase 10.1038/s41467-018-07090-4} {\bibfield  {journal} {\bibinfo  {journal} {Nature Communications}\ }\textbf {\bibinfo {volume} {9}},\ \bibinfo {pages} {4812} (\bibinfo {year} {2018})}\BibitemShut {NoStop}%
\bibitem [{\citenamefont {Wang}\ \emph {et~al.}(2021)\citenamefont {Wang}, \citenamefont {Fontana}, \citenamefont {Cerezo}, \citenamefont {Sharma}, \citenamefont {Sone}, \citenamefont {Cincio},\ and\ \citenamefont {Coles}}]{Wang2021noise}%
  \BibitemOpen
  \bibfield  {author} {\bibinfo {author} {\bibfnamefont {S.}~\bibnamefont {Wang}}, \bibinfo {author} {\bibfnamefont {E.}~\bibnamefont {Fontana}}, \bibinfo {author} {\bibfnamefont {M.}~\bibnamefont {Cerezo}}, \bibinfo {author} {\bibfnamefont {K.}~\bibnamefont {Sharma}}, \bibinfo {author} {\bibfnamefont {A.}~\bibnamefont {Sone}}, \bibinfo {author} {\bibfnamefont {L.}~\bibnamefont {Cincio}}, \ and\ \bibinfo {author} {\bibfnamefont {P.~J.}\ \bibnamefont {Coles}},\ }\href {\doibase 10.1038/s41467-021-27045-6} {\bibfield  {journal} {\bibinfo  {journal} {Nature Communications}\ }\textbf {\bibinfo {volume} {12}} (\bibinfo {year} {2021}),\ 10.1038/s41467-021-27045-6}\BibitemShut {NoStop}%
\bibitem [{\citenamefont {Ragone}\ \emph {et~al.}(2024)\citenamefont {Ragone}, \citenamefont {Bakalov}, \citenamefont {Sauvage}, \citenamefont {Kemper}, \citenamefont {Ortiz~Marrero}, \citenamefont {Larocca},\ and\ \citenamefont {Cerezo}}]{Ragone2024lie}%
  \BibitemOpen
  \bibfield  {author} {\bibinfo {author} {\bibfnamefont {M.}~\bibnamefont {Ragone}}, \bibinfo {author} {\bibfnamefont {B.~N.}\ \bibnamefont {Bakalov}}, \bibinfo {author} {\bibfnamefont {F.}~\bibnamefont {Sauvage}}, \bibinfo {author} {\bibfnamefont {A.~F.}\ \bibnamefont {Kemper}}, \bibinfo {author} {\bibfnamefont {C.}~\bibnamefont {Ortiz~Marrero}}, \bibinfo {author} {\bibfnamefont {M.}~\bibnamefont {Larocca}}, \ and\ \bibinfo {author} {\bibfnamefont {M.}~\bibnamefont {Cerezo}},\ }\href {\doibase 10.1038/s41467-024-49909-3} {\bibfield  {journal} {\bibinfo  {journal} {Nature Communications}\ }\textbf {\bibinfo {volume} {15}} (\bibinfo {year} {2024}),\ 10.1038/s41467-024-49909-3}\BibitemShut {NoStop}%
\bibitem [{\citenamefont {Jones}\ \emph {et~al.}(2024)\citenamefont {Jones}, \citenamefont {Kraus}, \citenamefont {Bhardwaj}, \citenamefont {Adler}, \citenamefont {Schrödl-Baumann},\ and\ \citenamefont {Manrique}}]{jones2024benchmarkingquantummodelstimeseries}%
  \BibitemOpen
  \bibfield  {author} {\bibinfo {author} {\bibfnamefont {C.}~\bibnamefont {Jones}}, \bibinfo {author} {\bibfnamefont {N.}~\bibnamefont {Kraus}}, \bibinfo {author} {\bibfnamefont {P.}~\bibnamefont {Bhardwaj}}, \bibinfo {author} {\bibfnamefont {M.}~\bibnamefont {Adler}}, \bibinfo {author} {\bibfnamefont {M.}~\bibnamefont {Schrödl-Baumann}}, \ and\ \bibinfo {author} {\bibfnamefont {D.~Z.}\ \bibnamefont {Manrique}},\ }\href {https://arxiv.org/abs/2412.13878} {\enquote {\bibinfo {title} {Benchmarking quantum models for time-series forecasting},}\ } (\bibinfo {year} {2024}),\ \Eprint {http://arxiv.org/abs/2412.13878} {arXiv:2412.13878 [quant-ph]} \BibitemShut {NoStop}%
\bibitem [{\citenamefont {Elliott}\ \emph {et~al.}(2022)\citenamefont {Elliott}, \citenamefont {Gu}, \citenamefont {Garner},\ and\ \citenamefont {Thompson}}]{elliottQuantumAdaptiveAgents2022}%
  \BibitemOpen
  \bibfield  {author} {\bibinfo {author} {\bibfnamefont {T.~J.}\ \bibnamefont {Elliott}}, \bibinfo {author} {\bibfnamefont {M.}~\bibnamefont {Gu}}, \bibinfo {author} {\bibfnamefont {A.~J.~P.}\ \bibnamefont {Garner}}, \ and\ \bibinfo {author} {\bibfnamefont {J.}~\bibnamefont {Thompson}},\ }\href {\doibase 10.1103/PhysRevX.12.011007} {\bibfield  {journal} {\bibinfo  {journal} {Physical Review X}\ }\textbf {\bibinfo {volume} {12}},\ \bibinfo {pages} {011007} (\bibinfo {year} {2022})}\BibitemShut {NoStop}%
\bibitem [{\citenamefont {Jaeger}\ and\ \citenamefont {Haas}(2004)}]{jaeger2004harnessing}%
  \BibitemOpen
  \bibfield  {author} {\bibinfo {author} {\bibfnamefont {H.}~\bibnamefont {Jaeger}}\ and\ \bibinfo {author} {\bibfnamefont {H.}~\bibnamefont {Haas}},\ }\href {\doibase 10.1126/science.1091277} {\bibfield  {journal} {\bibinfo  {journal} {science}\ }\textbf {\bibinfo {volume} {304}},\ \bibinfo {pages} {78} (\bibinfo {year} {2004})}\BibitemShut {NoStop}%
\bibitem [{\citenamefont {Maass}\ \emph {et~al.}(2002)\citenamefont {Maass}, \citenamefont {Natschl{\"a}ger},\ and\ \citenamefont {Markram}}]{maass2002real}%
  \BibitemOpen
  \bibfield  {author} {\bibinfo {author} {\bibfnamefont {W.}~\bibnamefont {Maass}}, \bibinfo {author} {\bibfnamefont {T.}~\bibnamefont {Natschl{\"a}ger}}, \ and\ \bibinfo {author} {\bibfnamefont {H.}~\bibnamefont {Markram}},\ }\href {\doibase 10.1162/089976602760407955} {\bibfield  {journal} {\bibinfo  {journal} {Neural computation}\ }\textbf {\bibinfo {volume} {14}},\ \bibinfo {pages} {2531} (\bibinfo {year} {2002})}\BibitemShut {NoStop}%
\bibitem [{\citenamefont {Fujii}\ and\ \citenamefont {Nakajima}(2017)}]{fujii2017harnessing}%
  \BibitemOpen
  \bibfield  {author} {\bibinfo {author} {\bibfnamefont {K.}~\bibnamefont {Fujii}}\ and\ \bibinfo {author} {\bibfnamefont {K.}~\bibnamefont {Nakajima}},\ }\href {\doibase 10.1103/PhysRevApplied.8.024030} {\bibfield  {journal} {\bibinfo  {journal} {Phys. Rev. Appl.}\ }\textbf {\bibinfo {volume} {8}},\ \bibinfo {pages} {024030} (\bibinfo {year} {2017})}\BibitemShut {NoStop}%
\bibitem [{\citenamefont {Fujii}\ and\ \citenamefont {Nakajima}(2021)}]{Fujii2021}%
  \BibitemOpen
  \bibfield  {author} {\bibinfo {author} {\bibfnamefont {K.}~\bibnamefont {Fujii}}\ and\ \bibinfo {author} {\bibfnamefont {K.}~\bibnamefont {Nakajima}},\ }in\ \href {\doibase 10.1007/978-981-13-1687-6_18} {\emph {\bibinfo {booktitle} {Reservoir Computing: Theory, Physical Implementations, and Applications}}},\ \bibinfo {editor} {edited by\ \bibinfo {editor} {\bibfnamefont {K.}~\bibnamefont {Nakajima}}\ and\ \bibinfo {editor} {\bibfnamefont {I.}~\bibnamefont {Fischer}}}\ (\bibinfo  {publisher} {Springer Singapore},\ \bibinfo {address} {Singapore},\ \bibinfo {year} {2021})\ pp.\ \bibinfo {pages} {423--450}\BibitemShut {NoStop}%
\bibitem [{\citenamefont {Innocenti}\ \emph {et~al.}(2023)\citenamefont {Innocenti}, \citenamefont {Lorenzo}, \citenamefont {Palmisano}, \citenamefont {Ferraro}, \citenamefont {Paternostro},\ and\ \citenamefont {Palma}}]{innocentiPotentialLimitationsQuantum2023a}%
  \BibitemOpen
  \bibfield  {author} {\bibinfo {author} {\bibfnamefont {L.}~\bibnamefont {Innocenti}}, \bibinfo {author} {\bibfnamefont {S.}~\bibnamefont {Lorenzo}}, \bibinfo {author} {\bibfnamefont {I.}~\bibnamefont {Palmisano}}, \bibinfo {author} {\bibfnamefont {A.}~\bibnamefont {Ferraro}}, \bibinfo {author} {\bibfnamefont {M.}~\bibnamefont {Paternostro}}, \ and\ \bibinfo {author} {\bibfnamefont {G.~M.}\ \bibnamefont {Palma}},\ }\href {\doibase 10.1038/s42005-023-01233-w} {\bibfield  {journal} {\bibinfo  {journal} {Communications Physics}\ }\textbf {\bibinfo {volume} {6}},\ \bibinfo {pages} {1} (\bibinfo {year} {2023})}\BibitemShut {NoStop}%
\bibitem [{\citenamefont {Chen}\ and\ \citenamefont {Nurdin}(2019)}]{chenLearningNonlinearInput2019}%
  \BibitemOpen
  \bibfield  {author} {\bibinfo {author} {\bibfnamefont {J.}~\bibnamefont {Chen}}\ and\ \bibinfo {author} {\bibfnamefont {H.~I.}\ \bibnamefont {Nurdin}},\ }\href {\doibase 10.1007/s11128-019-2311-9} {\bibfield  {journal} {\bibinfo  {journal} {Quantum Information Processing}\ }\textbf {\bibinfo {volume} {18}},\ \bibinfo {pages} {198} (\bibinfo {year} {2019})}\BibitemShut {NoStop}%
\bibitem [{\citenamefont {Xia}\ \emph {et~al.}(2023)\citenamefont {Xia}, \citenamefont {Zou}, \citenamefont {Qiu}, \citenamefont {Chen}, \citenamefont {Zhu}, \citenamefont {Li}, \citenamefont {Deng},\ and\ \citenamefont {Li}}]{xiaConfiguredQuantumReservoir2023}%
  \BibitemOpen
  \bibfield  {author} {\bibinfo {author} {\bibfnamefont {W.}~\bibnamefont {Xia}}, \bibinfo {author} {\bibfnamefont {J.}~\bibnamefont {Zou}}, \bibinfo {author} {\bibfnamefont {X.}~\bibnamefont {Qiu}}, \bibinfo {author} {\bibfnamefont {F.}~\bibnamefont {Chen}}, \bibinfo {author} {\bibfnamefont {B.}~\bibnamefont {Zhu}}, \bibinfo {author} {\bibfnamefont {C.}~\bibnamefont {Li}}, \bibinfo {author} {\bibfnamefont {D.-L.}\ \bibnamefont {Deng}}, \ and\ \bibinfo {author} {\bibfnamefont {X.}~\bibnamefont {Li}},\ }\href {\doibase 10.1016/j.scib.2023.08.040} {\bibfield  {journal} {\bibinfo  {journal} {Science Bulletin}\ }\textbf {\bibinfo {volume} {68}},\ \bibinfo {pages} {2321} (\bibinfo {year} {2023})}\BibitemShut {NoStop}%
\bibitem [{\citenamefont {Li}\ \emph {et~al.}(2025)\citenamefont {Li}, \citenamefont {Mukhopadhyay}, \citenamefont {Bayat},\ and\ \citenamefont {Habibnia}}]{li2025quantumreservoircomputingrealized}%
  \BibitemOpen
  \bibfield  {author} {\bibinfo {author} {\bibfnamefont {Q.}~\bibnamefont {Li}}, \bibinfo {author} {\bibfnamefont {C.}~\bibnamefont {Mukhopadhyay}}, \bibinfo {author} {\bibfnamefont {A.}~\bibnamefont {Bayat}}, \ and\ \bibinfo {author} {\bibfnamefont {A.}~\bibnamefont {Habibnia}},\ }\href {https://arxiv.org/abs/2505.13933} {\enquote {\bibinfo {title} {Quantum reservoir computing for realized volatility forecasting},}\ } (\bibinfo {year} {2025}),\ \Eprint {http://arxiv.org/abs/2505.13933} {arXiv:2505.13933 [quant-ph]} \BibitemShut {NoStop}%
\bibitem [{\citenamefont {Ghosh}\ \emph {et~al.}(2019{\natexlab{a}})\citenamefont {Ghosh}, \citenamefont {Opala}, \citenamefont {Matuszewski}, \citenamefont {Paterek},\ and\ \citenamefont {Liew}}]{ghoshQuantumReservoirProcessing2019}%
  \BibitemOpen
  \bibfield  {author} {\bibinfo {author} {\bibfnamefont {S.}~\bibnamefont {Ghosh}}, \bibinfo {author} {\bibfnamefont {A.}~\bibnamefont {Opala}}, \bibinfo {author} {\bibfnamefont {M.}~\bibnamefont {Matuszewski}}, \bibinfo {author} {\bibfnamefont {T.}~\bibnamefont {Paterek}}, \ and\ \bibinfo {author} {\bibfnamefont {T.~C.~H.}\ \bibnamefont {Liew}},\ }\href {\doibase 10.1038/s41534-019-0149-8} {\bibfield  {journal} {\bibinfo  {journal} {npj Quantum Information}\ }\textbf {\bibinfo {volume} {5}},\ \bibinfo {pages} {35} (\bibinfo {year} {2019}{\natexlab{a}})}\BibitemShut {NoStop}%
\bibitem [{\citenamefont {Angelatos}\ \emph {et~al.}(2021)\citenamefont {Angelatos}, \citenamefont {Khan},\ and\ \citenamefont {T{\"u}reci}}]{angelatosReservoirComputingApproach2021}%
  \BibitemOpen
  \bibfield  {author} {\bibinfo {author} {\bibfnamefont {G.}~\bibnamefont {Angelatos}}, \bibinfo {author} {\bibfnamefont {S.~A.}\ \bibnamefont {Khan}}, \ and\ \bibinfo {author} {\bibfnamefont {H.~E.}\ \bibnamefont {T{\"u}reci}},\ }\href {\doibase 10.1103/PhysRevX.11.041062} {\bibfield  {journal} {\bibinfo  {journal} {Physical Review X}\ }\textbf {\bibinfo {volume} {11}},\ \bibinfo {pages} {041062} (\bibinfo {year} {2021})}\BibitemShut {NoStop}%
\bibitem [{\citenamefont {Krisnanda}\ \emph {et~al.}(2023)\citenamefont {Krisnanda}, \citenamefont {Xu}, \citenamefont {Ghosh},\ and\ \citenamefont {Liew}}]{krisnandaTomographicCompletenessRobustness2023}%
  \BibitemOpen
  \bibfield  {author} {\bibinfo {author} {\bibfnamefont {T.}~\bibnamefont {Krisnanda}}, \bibinfo {author} {\bibfnamefont {H.}~\bibnamefont {Xu}}, \bibinfo {author} {\bibfnamefont {S.}~\bibnamefont {Ghosh}}, \ and\ \bibinfo {author} {\bibfnamefont {T.~C.~H.}\ \bibnamefont {Liew}},\ }\href {\doibase 10.1103/PhysRevA.107.042402} {\bibfield  {journal} {\bibinfo  {journal} {Physical Review A}\ }\textbf {\bibinfo {volume} {107}},\ \bibinfo {pages} {042402} (\bibinfo {year} {2023})}\BibitemShut {NoStop}%
\bibitem [{\citenamefont {Krisnanda}\ \emph {et~al.}(2022)\citenamefont {Krisnanda}, \citenamefont {Ghosh}, \citenamefont {Paterek}, \citenamefont {Laskowski},\ and\ \citenamefont {Liew}}]{krisnandaPhaseMeasurementStandard2022}%
  \BibitemOpen
  \bibfield  {author} {\bibinfo {author} {\bibfnamefont {T.}~\bibnamefont {Krisnanda}}, \bibinfo {author} {\bibfnamefont {S.}~\bibnamefont {Ghosh}}, \bibinfo {author} {\bibfnamefont {T.}~\bibnamefont {Paterek}}, \bibinfo {author} {\bibfnamefont {W.}~\bibnamefont {Laskowski}}, \ and\ \bibinfo {author} {\bibfnamefont {T.~C.}\ \bibnamefont {Liew}},\ }\href {\doibase 10.1103/PhysRevApplied.18.034011} {\bibfield  {journal} {\bibinfo  {journal} {Physical Review Applied}\ }\textbf {\bibinfo {volume} {18}},\ \bibinfo {pages} {034011} (\bibinfo {year} {2022})}\BibitemShut {NoStop}%
\bibitem [{\citenamefont {Li}\ \emph {et~al.}(2024)\citenamefont {Li}, \citenamefont {Ghosh}, \citenamefont {Shang}, \citenamefont {Xiong},\ and\ \citenamefont {Zhang}}]{liEstimatingManyProperties2024}%
  \BibitemOpen
  \bibfield  {author} {\bibinfo {author} {\bibfnamefont {Y.}~\bibnamefont {Li}}, \bibinfo {author} {\bibfnamefont {S.}~\bibnamefont {Ghosh}}, \bibinfo {author} {\bibfnamefont {J.}~\bibnamefont {Shang}}, \bibinfo {author} {\bibfnamefont {Q.}~\bibnamefont {Xiong}}, \ and\ \bibinfo {author} {\bibfnamefont {X.}~\bibnamefont {Zhang}},\ }\href {\doibase 10.1103/PhysRevResearch.6.013211} {\bibfield  {journal} {\bibinfo  {journal} {Physical Review Research}\ }\textbf {\bibinfo {volume} {6}},\ \bibinfo {pages} {013211} (\bibinfo {year} {2024})}\BibitemShut {NoStop}%
\bibitem [{\citenamefont {Suprano}\ \emph {et~al.}(2024)\citenamefont {Suprano}, \citenamefont {Zia}, \citenamefont {Innocenti}, \citenamefont {Lorenzo}, \citenamefont {Cimini}, \citenamefont {Giordani}, \citenamefont {Palmisano}, \citenamefont {Polino}, \citenamefont {Spagnolo}, \citenamefont {Sciarrino}, \citenamefont {Palma}, \citenamefont {Ferraro},\ and\ \citenamefont {Paternostro}}]{supranoExperimentalPropertyReconstruction2024}%
  \BibitemOpen
  \bibfield  {author} {\bibinfo {author} {\bibfnamefont {A.}~\bibnamefont {Suprano}}, \bibinfo {author} {\bibfnamefont {D.}~\bibnamefont {Zia}}, \bibinfo {author} {\bibfnamefont {L.}~\bibnamefont {Innocenti}}, \bibinfo {author} {\bibfnamefont {S.}~\bibnamefont {Lorenzo}}, \bibinfo {author} {\bibfnamefont {V.}~\bibnamefont {Cimini}}, \bibinfo {author} {\bibfnamefont {T.}~\bibnamefont {Giordani}}, \bibinfo {author} {\bibfnamefont {I.}~\bibnamefont {Palmisano}}, \bibinfo {author} {\bibfnamefont {E.}~\bibnamefont {Polino}}, \bibinfo {author} {\bibfnamefont {N.}~\bibnamefont {Spagnolo}}, \bibinfo {author} {\bibfnamefont {F.}~\bibnamefont {Sciarrino}}, \bibinfo {author} {\bibfnamefont {G.~M.}\ \bibnamefont {Palma}}, \bibinfo {author} {\bibfnamefont {A.}~\bibnamefont {Ferraro}}, \ and\ \bibinfo {author} {\bibfnamefont {M.}~\bibnamefont {Paternostro}},\ }\href {\doibase 10.1103/PhysRevLett.132.160802} {\bibfield  {journal} {\bibinfo  {journal} {Physical Review Letters}\ }\textbf {\bibinfo {volume} {132}},\ \bibinfo
  {pages} {160802} (\bibinfo {year} {2024})}\BibitemShut {NoStop}%
\bibitem [{\citenamefont {Ghosh}\ \emph {et~al.}(2019{\natexlab{b}})\citenamefont {Ghosh}, \citenamefont {Paterek},\ and\ \citenamefont {Liew}}]{ghoshQuantumNeuromorphicPlatform2019}%
  \BibitemOpen
  \bibfield  {author} {\bibinfo {author} {\bibfnamefont {S.}~\bibnamefont {Ghosh}}, \bibinfo {author} {\bibfnamefont {T.}~\bibnamefont {Paterek}}, \ and\ \bibinfo {author} {\bibfnamefont {T.~C.~H.}\ \bibnamefont {Liew}},\ }\href {\doibase 10.1103/PhysRevLett.123.260404} {\bibfield  {journal} {\bibinfo  {journal} {Physical Review Letters}\ }\textbf {\bibinfo {volume} {123}},\ \bibinfo {pages} {260404} (\bibinfo {year} {2019}{\natexlab{b}})}\BibitemShut {NoStop}%
\bibitem [{\citenamefont {Zhu}\ \emph {et~al.}(2025)\citenamefont {Zhu}, \citenamefont {Ehlers}, \citenamefont {Nurdin},\ and\ \citenamefont {Soh}}]{ZhuPractical2025}%
  \BibitemOpen
  \bibfield  {author} {\bibinfo {author} {\bibfnamefont {C.}~\bibnamefont {Zhu}}, \bibinfo {author} {\bibfnamefont {P.~J.}\ \bibnamefont {Ehlers}}, \bibinfo {author} {\bibfnamefont {H.~I.}\ \bibnamefont {Nurdin}}, \ and\ \bibinfo {author} {\bibfnamefont {D.}~\bibnamefont {Soh}},\ }\href {\doibase 10.1103/wsyq-jyxd} {\bibfield  {journal} {\bibinfo  {journal} {Phys. Rev. Res.}\ }\textbf {\bibinfo {volume} {7}},\ \bibinfo {pages} {023290} (\bibinfo {year} {2025})}\BibitemShut {NoStop}%
\bibitem [{\citenamefont {Dudas}\ \emph {et~al.}(2023)\citenamefont {Dudas}, \citenamefont {Carles}, \citenamefont {Plouet}, \citenamefont {Mizrahi}, \citenamefont {Grollier},\ and\ \citenamefont {Markovi{\'c}}}]{dudasQuantumReservoirComputing2023}%
  \BibitemOpen
  \bibfield  {author} {\bibinfo {author} {\bibfnamefont {J.}~\bibnamefont {Dudas}}, \bibinfo {author} {\bibfnamefont {B.}~\bibnamefont {Carles}}, \bibinfo {author} {\bibfnamefont {E.}~\bibnamefont {Plouet}}, \bibinfo {author} {\bibfnamefont {F.~A.}\ \bibnamefont {Mizrahi}}, \bibinfo {author} {\bibfnamefont {J.}~\bibnamefont {Grollier}}, \ and\ \bibinfo {author} {\bibfnamefont {D.}~\bibnamefont {Markovi{\'c}}},\ }\href {\doibase 10.1038/s41534-023-00734-4} {\bibfield  {journal} {\bibinfo  {journal} {npj Quantum Information}\ }\textbf {\bibinfo {volume} {9}},\ \bibinfo {pages} {64} (\bibinfo {year} {2023})}\BibitemShut {NoStop}%
\bibitem [{\citenamefont {Bravo}\ \emph {et~al.}(2022)\citenamefont {Bravo}, \citenamefont {Najafi}, \citenamefont {Gao},\ and\ \citenamefont {Yelin}}]{bravoQuantumReservoirComputing2022}%
  \BibitemOpen
  \bibfield  {author} {\bibinfo {author} {\bibfnamefont {R.~A.}\ \bibnamefont {Bravo}}, \bibinfo {author} {\bibfnamefont {K.}~\bibnamefont {Najafi}}, \bibinfo {author} {\bibfnamefont {X.}~\bibnamefont {Gao}}, \ and\ \bibinfo {author} {\bibfnamefont {S.~F.}\ \bibnamefont {Yelin}},\ }\href {\doibase 10.1103/PRXQuantum.3.030325} {\bibfield  {journal} {\bibinfo  {journal} {PRX Quantum}\ }\textbf {\bibinfo {volume} {3}},\ \bibinfo {pages} {030325} (\bibinfo {year} {2022})}\BibitemShut {NoStop}%
\bibitem [{\citenamefont {Negoro}\ \emph {et~al.}(2018)\citenamefont {Negoro}, \citenamefont {Mitarai}, \citenamefont {Fujii}, \citenamefont {Nakajima},\ and\ \citenamefont {Kitagawa}}]{negoroMachineLearningControllable2018}%
  \BibitemOpen
  \bibfield  {author} {\bibinfo {author} {\bibfnamefont {M.}~\bibnamefont {Negoro}}, \bibinfo {author} {\bibfnamefont {K.}~\bibnamefont {Mitarai}}, \bibinfo {author} {\bibfnamefont {K.}~\bibnamefont {Fujii}}, \bibinfo {author} {\bibfnamefont {K.}~\bibnamefont {Nakajima}}, \ and\ \bibinfo {author} {\bibfnamefont {M.}~\bibnamefont {Kitagawa}},\ }\href {\doibase 10.48550/arXiv.1806.10910} {\bibfield  {journal} {\bibinfo  {journal} {arXiv preprint arXiv:1806.10910}\ } (\bibinfo {year} {2018}),\ 10.48550/arXiv.1806.10910},\ \Eprint {http://arxiv.org/abs/1806.10910} {arXiv:1806.10910 [quant-ph]} \BibitemShut {NoStop}%
\bibitem [{\citenamefont {Chen}\ \emph {et~al.}(2020)\citenamefont {Chen}, \citenamefont {Nurdin},\ and\ \citenamefont {Yamamoto}}]{chenTemporalInformationProcessing2020}%
  \BibitemOpen
  \bibfield  {author} {\bibinfo {author} {\bibfnamefont {J.}~\bibnamefont {Chen}}, \bibinfo {author} {\bibfnamefont {H.~I.}\ \bibnamefont {Nurdin}}, \ and\ \bibinfo {author} {\bibfnamefont {N.}~\bibnamefont {Yamamoto}},\ }\href {\doibase 10.1103/PhysRevApplied.14.024065} {\bibfield  {journal} {\bibinfo  {journal} {Physical Review Applied}\ }\textbf {\bibinfo {volume} {14}},\ \bibinfo {pages} {024065} (\bibinfo {year} {2020})}\BibitemShut {NoStop}%
\bibitem [{\citenamefont {Tran}\ and\ \citenamefont {Nakajima}(2021)}]{tranLearningTemporalQuantum2021}%
  \BibitemOpen
  \bibfield  {author} {\bibinfo {author} {\bibfnamefont {Q.~H.}\ \bibnamefont {Tran}}\ and\ \bibinfo {author} {\bibfnamefont {K.}~\bibnamefont {Nakajima}},\ }\href {\doibase 10.1103/PhysRevLett.127.260401} {\bibfield  {journal} {\bibinfo  {journal} {Physical Review Letters}\ }\textbf {\bibinfo {volume} {127}},\ \bibinfo {pages} {260401} (\bibinfo {year} {2021})}\BibitemShut {NoStop}%
\bibitem [{\citenamefont {Zhang}\ \emph {et~al.}(2021)\citenamefont {Zhang}, \citenamefont {Fan}, \citenamefont {Wang},\ and\ \citenamefont {Wang}}]{zhangLearningHamiltonianDynamics2021}%
  \BibitemOpen
  \bibfield  {author} {\bibinfo {author} {\bibfnamefont {H.}~\bibnamefont {Zhang}}, \bibinfo {author} {\bibfnamefont {H.}~\bibnamefont {Fan}}, \bibinfo {author} {\bibfnamefont {L.}~\bibnamefont {Wang}}, \ and\ \bibinfo {author} {\bibfnamefont {X.}~\bibnamefont {Wang}},\ }\href {\doibase 10.1103/PhysRevE.104.024205} {\bibfield  {journal} {\bibinfo  {journal} {Physical Review E}\ }\textbf {\bibinfo {volume} {104}},\ \bibinfo {pages} {024205} (\bibinfo {year} {2021})}\BibitemShut {NoStop}%
\bibitem [{\citenamefont {Suzuki}\ \emph {et~al.}(2022)\citenamefont {Suzuki}, \citenamefont {Gao}, \citenamefont {Pradel}, \citenamefont {Yasuoka},\ and\ \citenamefont {Yamamoto}}]{suzukiNaturalQuantumReservoir2022}%
  \BibitemOpen
  \bibfield  {author} {\bibinfo {author} {\bibfnamefont {Y.}~\bibnamefont {Suzuki}}, \bibinfo {author} {\bibfnamefont {Q.}~\bibnamefont {Gao}}, \bibinfo {author} {\bibfnamefont {K.~C.}\ \bibnamefont {Pradel}}, \bibinfo {author} {\bibfnamefont {K.}~\bibnamefont {Yasuoka}}, \ and\ \bibinfo {author} {\bibfnamefont {N.}~\bibnamefont {Yamamoto}},\ }\href {\doibase 10.1038/s41598-022-05061-w} {\bibfield  {journal} {\bibinfo  {journal} {Scientific Reports}\ }\textbf {\bibinfo {volume} {12}},\ \bibinfo {pages} {1353} (\bibinfo {year} {2022})}\BibitemShut {NoStop}%
\bibitem [{\citenamefont {Mujal}\ \emph {et~al.}(2023{\natexlab{a}})\citenamefont {Mujal}, \citenamefont {{Mart{\'i}nez-Pe{\~n}a}}, \citenamefont {Giorgi}, \citenamefont {Soriano},\ and\ \citenamefont {Zambrini}}]{mujalTimeseriesQuantumReservoir2023}%
  \BibitemOpen
  \bibfield  {author} {\bibinfo {author} {\bibfnamefont {P.}~\bibnamefont {Mujal}}, \bibinfo {author} {\bibfnamefont {R.}~\bibnamefont {{Mart{\'i}nez-Pe{\~n}a}}}, \bibinfo {author} {\bibfnamefont {G.~L.}\ \bibnamefont {Giorgi}}, \bibinfo {author} {\bibfnamefont {M.~C.}\ \bibnamefont {Soriano}}, \ and\ \bibinfo {author} {\bibfnamefont {R.}~\bibnamefont {Zambrini}},\ }\href {\doibase 10.1038/s41534-023-00682-z} {\bibfield  {journal} {\bibinfo  {journal} {npj Quantum Information}\ }\textbf {\bibinfo {volume} {9}},\ \bibinfo {pages} {16} (\bibinfo {year} {2023}{\natexlab{a}})}\BibitemShut {NoStop}%
\bibitem [{\citenamefont {Kobayashi}\ \emph {et~al.}(2024)\citenamefont {Kobayashi}, \citenamefont {Fujii},\ and\ \citenamefont {Yamamoto}}]{kobayashiFeedbackDrivenQuantumReservoir2024}%
  \BibitemOpen
  \bibfield  {author} {\bibinfo {author} {\bibfnamefont {K.}~\bibnamefont {Kobayashi}}, \bibinfo {author} {\bibfnamefont {K.}~\bibnamefont {Fujii}}, \ and\ \bibinfo {author} {\bibfnamefont {N.}~\bibnamefont {Yamamoto}},\ }\href {\doibase 10.1103/PRXQuantum.5.040325} {\bibfield  {journal} {\bibinfo  {journal} {PRX Quantum}\ }\textbf {\bibinfo {volume} {5}},\ \bibinfo {pages} {040325} (\bibinfo {year} {2024})}\BibitemShut {NoStop}%
\bibitem [{\citenamefont {Salatino}\ \emph {et~al.}(2025)\citenamefont {Salatino}, \citenamefont {Mariani}, \citenamefont {Giordano}, \citenamefont {D'Amore}, \citenamefont {Mastroianni}, \citenamefont {Pontieri}, \citenamefont {Vinci}, \citenamefont {Gencarelli}, \citenamefont {Primavera}, \citenamefont {Plastina}, \citenamefont {Settino},\ and\ \citenamefont {Carbone}}]{salatino2025forecastinglowdimensionalturbulencemultidimensional}%
  \BibitemOpen
  \bibfield  {author} {\bibinfo {author} {\bibfnamefont {L.}~\bibnamefont {Salatino}}, \bibinfo {author} {\bibfnamefont {L.}~\bibnamefont {Mariani}}, \bibinfo {author} {\bibfnamefont {A.}~\bibnamefont {Giordano}}, \bibinfo {author} {\bibfnamefont {F.}~\bibnamefont {D'Amore}}, \bibinfo {author} {\bibfnamefont {C.}~\bibnamefont {Mastroianni}}, \bibinfo {author} {\bibfnamefont {L.}~\bibnamefont {Pontieri}}, \bibinfo {author} {\bibfnamefont {A.}~\bibnamefont {Vinci}}, \bibinfo {author} {\bibfnamefont {C.}~\bibnamefont {Gencarelli}}, \bibinfo {author} {\bibfnamefont {L.}~\bibnamefont {Primavera}}, \bibinfo {author} {\bibfnamefont {F.}~\bibnamefont {Plastina}}, \bibinfo {author} {\bibfnamefont {J.}~\bibnamefont {Settino}}, \ and\ \bibinfo {author} {\bibfnamefont {F.}~\bibnamefont {Carbone}},\ }\href {https://arxiv.org/abs/2509.04006} {\enquote {\bibinfo {title} {Forecasting low-dimensional turbulence via multi-dimensional hybrid quantum reservoir computing},}\ } (\bibinfo {year} {2025}),\ \Eprint
  {http://arxiv.org/abs/2509.04006} {arXiv:2509.04006 [quant-ph]} \BibitemShut {NoStop}%
\bibitem [{\citenamefont {De~Lorenzis}\ \emph {et~al.}(2025)\citenamefont {De~Lorenzis}, \citenamefont {Casado}, \citenamefont {Estarellas}, \citenamefont {Lo~Gullo}, \citenamefont {Lux}, \citenamefont {Plastina}, \citenamefont {Riera},\ and\ \citenamefont {Settino}}]{delorenzis2025harnessing}%
  \BibitemOpen
  \bibfield  {author} {\bibinfo {author} {\bibfnamefont {A.}~\bibnamefont {De~Lorenzis}}, \bibinfo {author} {\bibfnamefont {M.}~\bibnamefont {Casado}}, \bibinfo {author} {\bibfnamefont {M.}~\bibnamefont {Estarellas}}, \bibinfo {author} {\bibfnamefont {N.}~\bibnamefont {Lo~Gullo}}, \bibinfo {author} {\bibfnamefont {T.}~\bibnamefont {Lux}}, \bibinfo {author} {\bibfnamefont {F.}~\bibnamefont {Plastina}}, \bibinfo {author} {\bibfnamefont {A.}~\bibnamefont {Riera}}, \ and\ \bibinfo {author} {\bibfnamefont {J.}~\bibnamefont {Settino}},\ }\href {\doibase 10.1103/PhysRevApplied.23.044024} {\bibfield  {journal} {\bibinfo  {journal} {Phys. Rev. Appl.}\ }\textbf {\bibinfo {volume} {23}},\ \bibinfo {pages} {044024} (\bibinfo {year} {2025})}\BibitemShut {NoStop}%
\bibitem [{\citenamefont {Hou}\ \emph {et~al.}(2025)\citenamefont {Hou}, \citenamefont {Hua}, \citenamefont {Wu}, \citenamefont {Xia}, \citenamefont {Chen}, \citenamefont {Li}, \citenamefont {Li}, \citenamefont {Peng},\ and\ \citenamefont {Du}}]{hou2025highaccuracy}%
  \BibitemOpen
  \bibfield  {author} {\bibinfo {author} {\bibfnamefont {Y.}~\bibnamefont {Hou}}, \bibinfo {author} {\bibfnamefont {J.}~\bibnamefont {Hua}}, \bibinfo {author} {\bibfnamefont {Z.}~\bibnamefont {Wu}}, \bibinfo {author} {\bibfnamefont {W.}~\bibnamefont {Xia}}, \bibinfo {author} {\bibfnamefont {Y.}~\bibnamefont {Chen}}, \bibinfo {author} {\bibfnamefont {X.}~\bibnamefont {Li}}, \bibinfo {author} {\bibfnamefont {Z.}~\bibnamefont {Li}}, \bibinfo {author} {\bibfnamefont {X.}~\bibnamefont {Peng}}, \ and\ \bibinfo {author} {\bibfnamefont {J.}~\bibnamefont {Du}},\ }\href@noop {} {\enquote {\bibinfo {title} {High-accuracy temporal prediction via experimental quantum reservoir computing in correlated spins},}\ } (\bibinfo {year} {2025}),\ \Eprint {http://arxiv.org/abs/2508.12383} {arXiv:2508.12383 [quant-ph]} \BibitemShut {NoStop}%
\bibitem [{\citenamefont {May}(1976)}]{maySimpleMathematicalModels1976}%
  \BibitemOpen
  \bibfield  {author} {\bibinfo {author} {\bibfnamefont {R.~M.}\ \bibnamefont {May}},\ }\href {\doibase 10.1038/261459a0} {\bibfield  {journal} {\bibinfo  {journal} {Nature}\ }\textbf {\bibinfo {volume} {261}},\ \bibinfo {pages} {459} (\bibinfo {year} {1976})}\BibitemShut {NoStop}%
\bibitem [{\citenamefont {Strogatz}(2019)}]{strogatzNonlinearDynamicsChaos2019}%
  \BibitemOpen
  \bibfield  {author} {\bibinfo {author} {\bibfnamefont {S.}~\bibnamefont {Strogatz}},\ }\href@noop {} {\emph {\bibinfo {title} {Nonlinear Dynamics and Chaos: With Applications to Physics, Biology, Chemistry, and Engineering}}},\ \bibinfo {edition} {second edition}\ ed.\ (\bibinfo  {publisher} {CRC Press},\ \bibinfo {address} {Boca Raton},\ \bibinfo {year} {2019})\BibitemShut {NoStop}%
\bibitem [{\citenamefont {H{\'e}non}(1976)}]{henonTwodimensionalMappingStrange1976}%
  \BibitemOpen
  \bibfield  {author} {\bibinfo {author} {\bibfnamefont {M.}~\bibnamefont {H{\'e}non}},\ }\href {\doibase 10.1007/BF01608556} {\bibfield  {journal} {\bibinfo  {journal} {Communications in Mathematical Physics}\ }\textbf {\bibinfo {volume} {50}},\ \bibinfo {pages} {69} (\bibinfo {year} {1976})}\BibitemShut {NoStop}%
\bibitem [{\citenamefont {Benedicks}\ and\ \citenamefont {Carleson}(1991)}]{benedicks1991dynamics}%
  \BibitemOpen
  \bibfield  {author} {\bibinfo {author} {\bibfnamefont {M.}~\bibnamefont {Benedicks}}\ and\ \bibinfo {author} {\bibfnamefont {L.}~\bibnamefont {Carleson}},\ }\href@noop {} {\bibfield  {journal} {\bibinfo  {journal} {Annals of Mathematics}\ }\textbf {\bibinfo {volume} {133}},\ \bibinfo {pages} {73} (\bibinfo {year} {1991})}\BibitemShut {NoStop}%
\bibitem [{\citenamefont {Ghosh}\ \emph {et~al.}(2019{\natexlab{c}})\citenamefont {Ghosh}, \citenamefont {Opala}, \citenamefont {Matuszewski}, \citenamefont {Paterek},\ and\ \citenamefont {Liew}}]{ghosh2019quantum}%
  \BibitemOpen
  \bibfield  {author} {\bibinfo {author} {\bibfnamefont {S.}~\bibnamefont {Ghosh}}, \bibinfo {author} {\bibfnamefont {A.}~\bibnamefont {Opala}}, \bibinfo {author} {\bibfnamefont {M.}~\bibnamefont {Matuszewski}}, \bibinfo {author} {\bibfnamefont {T.}~\bibnamefont {Paterek}}, \ and\ \bibinfo {author} {\bibfnamefont {T.~C.}\ \bibnamefont {Liew}},\ }\href {\doibase https://doi.org/10.1038/s41534-019-0149-8} {\bibfield  {journal} {\bibinfo  {journal} {npj Quantum Information}\ }\textbf {\bibinfo {volume} {5}},\ \bibinfo {pages} {35} (\bibinfo {year} {2019}{\natexlab{c}})}\BibitemShut {NoStop}%
\bibitem [{\citenamefont {Mujal}\ \emph {et~al.}(2021)\citenamefont {Mujal}, \citenamefont {Mart{\'\i}nez-Pe{\~n}a}, \citenamefont {Nokkala}, \citenamefont {Garc{\'\i}a-Beni}, \citenamefont {Giorgi}, \citenamefont {Soriano},\ and\ \citenamefont {Zambrini}}]{mujal2021opportunities}%
  \BibitemOpen
  \bibfield  {author} {\bibinfo {author} {\bibfnamefont {P.}~\bibnamefont {Mujal}}, \bibinfo {author} {\bibfnamefont {R.}~\bibnamefont {Mart{\'\i}nez-Pe{\~n}a}}, \bibinfo {author} {\bibfnamefont {J.}~\bibnamefont {Nokkala}}, \bibinfo {author} {\bibfnamefont {J.}~\bibnamefont {Garc{\'\i}a-Beni}}, \bibinfo {author} {\bibfnamefont {G.~L.}\ \bibnamefont {Giorgi}}, \bibinfo {author} {\bibfnamefont {M.~C.}\ \bibnamefont {Soriano}}, \ and\ \bibinfo {author} {\bibfnamefont {R.}~\bibnamefont {Zambrini}},\ }\href {\doibase https://doi.org/10.1002/qute.202100027} {\bibfield  {journal} {\bibinfo  {journal} {Advanced Quantum Technologies}\ }\textbf {\bibinfo {volume} {4}},\ \bibinfo {pages} {2100027} (\bibinfo {year} {2021})}\BibitemShut {NoStop}%
\bibitem [{\citenamefont {Mujal}\ \emph {et~al.}(2023{\natexlab{b}})\citenamefont {Mujal}, \citenamefont {Mart{\'\i}nez-Pe{\~n}a}, \citenamefont {Giorgi}, \citenamefont {Soriano},\ and\ \citenamefont {Zambrini}}]{mujal2023time}%
  \BibitemOpen
  \bibfield  {author} {\bibinfo {author} {\bibfnamefont {P.}~\bibnamefont {Mujal}}, \bibinfo {author} {\bibfnamefont {R.}~\bibnamefont {Mart{\'\i}nez-Pe{\~n}a}}, \bibinfo {author} {\bibfnamefont {G.~L.}\ \bibnamefont {Giorgi}}, \bibinfo {author} {\bibfnamefont {M.~C.}\ \bibnamefont {Soriano}}, \ and\ \bibinfo {author} {\bibfnamefont {R.}~\bibnamefont {Zambrini}},\ }\href {\doibase https://doi.org/10.1038/s41534-023-00682-z} {\bibfield  {journal} {\bibinfo  {journal} {npj Quantum Information}\ }\textbf {\bibinfo {volume} {9}},\ \bibinfo {pages} {16} (\bibinfo {year} {2023}{\natexlab{b}})}\BibitemShut {NoStop}%
\bibitem [{\citenamefont {Garcia-Beni}\ \emph {et~al.}(2024)\citenamefont {Garcia-Beni}, \citenamefont {Giorgi}, \citenamefont {Soriano},\ and\ \citenamefont {Zambrini}}]{garcia2024quantum}%
  \BibitemOpen
  \bibfield  {author} {\bibinfo {author} {\bibfnamefont {J.}~\bibnamefont {Garcia-Beni}}, \bibinfo {author} {\bibfnamefont {G.~L.}\ \bibnamefont {Giorgi}}, \bibinfo {author} {\bibfnamefont {M.~C.}\ \bibnamefont {Soriano}}, \ and\ \bibinfo {author} {\bibfnamefont {R.}~\bibnamefont {Zambrini}},\ }in\ \href {\doibase 10.1117/12.3027999} {\emph {\bibinfo {booktitle} {Quantum Communications and Quantum Imaging XXII}}},\ Vol.\ \bibinfo {volume} {PC13148},\ \bibinfo {editor} {edited by\ \bibinfo {editor} {\bibfnamefont {K.~S.}\ \bibnamefont {Deacon}}\ and\ \bibinfo {editor} {\bibfnamefont {R.~E.}\ \bibnamefont {Meyers}}},\ \bibinfo {organization} {International Society for Optics and Photonics}\ (\bibinfo  {publisher} {SPIE},\ \bibinfo {year} {2024})\ p.\ \bibinfo {pages} {PC131480E}\BibitemShut {NoStop}%
\bibitem [{\citenamefont {Mitarai}\ \emph {et~al.}(2018)\citenamefont {Mitarai}, \citenamefont {Negoro}, \citenamefont {Kitagawa},\ and\ \citenamefont {Fujii}}]{mitaraiQuantumCircuitLearning2018}%
  \BibitemOpen
  \bibfield  {author} {\bibinfo {author} {\bibfnamefont {K.}~\bibnamefont {Mitarai}}, \bibinfo {author} {\bibfnamefont {M.}~\bibnamefont {Negoro}}, \bibinfo {author} {\bibfnamefont {M.}~\bibnamefont {Kitagawa}}, \ and\ \bibinfo {author} {\bibfnamefont {K.}~\bibnamefont {Fujii}},\ }\href {\doibase 10.1103/PhysRevA.98.032309} {\bibfield  {journal} {\bibinfo  {journal} {Physical Review A}\ }\textbf {\bibinfo {volume} {98}},\ \bibinfo {pages} {032309} (\bibinfo {year} {2018})}\BibitemShut {NoStop}%
\bibitem [{\citenamefont {Schuld}\ \emph {et~al.}(2021)\citenamefont {Schuld}, \citenamefont {Sweke},\ and\ \citenamefont {Meyer}}]{schuldEffectDataEncoding2021}%
  \BibitemOpen
  \bibfield  {author} {\bibinfo {author} {\bibfnamefont {M.}~\bibnamefont {Schuld}}, \bibinfo {author} {\bibfnamefont {R.}~\bibnamefont {Sweke}}, \ and\ \bibinfo {author} {\bibfnamefont {J.~J.}\ \bibnamefont {Meyer}},\ }\href {\doibase 10.1103/PhysRevA.103.032430} {\bibfield  {journal} {\bibinfo  {journal} {Physical Review A}\ }\textbf {\bibinfo {volume} {103}},\ \bibinfo {pages} {032430} (\bibinfo {year} {2021})}\BibitemShut {NoStop}%
\bibitem [{\citenamefont {Nielsen}\ and\ \citenamefont {Chuang}(2012)}]{nielsen2010quantum}%
  \BibitemOpen
  \bibfield  {author} {\bibinfo {author} {\bibfnamefont {M.~A.}\ \bibnamefont {Nielsen}}\ and\ \bibinfo {author} {\bibfnamefont {I.~L.}\ \bibnamefont {Chuang}},\ }\href@noop {} {\emph {\bibinfo {title} {Quantum Computation and Quantum Information}}},\ \bibinfo {edition} {10th}\ ed.\ (\bibinfo  {publisher} {Cambridge University Press},\ \bibinfo {address} {Cambridge},\ \bibinfo {year} {2012})\BibitemShut {NoStop}%
\bibitem [{\citenamefont {Dubey}\ \emph {et~al.}(2012)\citenamefont {Dubey}, \citenamefont {Silvestri}, \citenamefont {Finn}, \citenamefont {Vinjanampathy},\ and\ \citenamefont {Jacobs}}]{dubey2012approach}%
  \BibitemOpen
  \bibfield  {author} {\bibinfo {author} {\bibfnamefont {S.}~\bibnamefont {Dubey}}, \bibinfo {author} {\bibfnamefont {L.}~\bibnamefont {Silvestri}}, \bibinfo {author} {\bibfnamefont {J.}~\bibnamefont {Finn}}, \bibinfo {author} {\bibfnamefont {S.}~\bibnamefont {Vinjanampathy}}, \ and\ \bibinfo {author} {\bibfnamefont {K.}~\bibnamefont {Jacobs}},\ }\href {\doibase https://doi.org/10.1103/PhysRevE.85.011141} {\bibfield  {journal} {\bibinfo  {journal} {Phys. Rev. E}\ }\textbf {\bibinfo {volume} {85}},\ \bibinfo {pages} {011141} (\bibinfo {year} {2012})}\BibitemShut {NoStop}%
\bibitem [{\citenamefont {Ithier}\ and\ \citenamefont {Benaych-Georges}(2017)}]{ithier2017dynamical}%
  \BibitemOpen
  \bibfield  {author} {\bibinfo {author} {\bibfnamefont {G.}~\bibnamefont {Ithier}}\ and\ \bibinfo {author} {\bibfnamefont {F.}~\bibnamefont {Benaych-Georges}},\ }\href {\doibase 10.1103/PhysRevA.96.012108} {\bibfield  {journal} {\bibinfo  {journal} {Phys. Rev. A}\ }\textbf {\bibinfo {volume} {96}},\ \bibinfo {pages} {012108} (\bibinfo {year} {2017})}\BibitemShut {NoStop}%
\bibitem [{\citenamefont {Reimann}(2018)}]{reimann2018dynamical}%
  \BibitemOpen
  \bibfield  {author} {\bibinfo {author} {\bibfnamefont {P.}~\bibnamefont {Reimann}},\ }\href {\doibase https://doi.org/10.1103/PhysRevE.97.062129} {\bibfield  {journal} {\bibinfo  {journal} {Phys. Rev. E}\ }\textbf {\bibinfo {volume} {97}},\ \bibinfo {pages} {062129} (\bibinfo {year} {2018})}\BibitemShut {NoStop}%
\bibitem [{\citenamefont {Ahmed}\ \emph {et~al.}(2025{\natexlab{a}})\citenamefont {Ahmed}, \citenamefont {Tennie},\ and\ \citenamefont {Magri}}]{ahmedOptimalTrainingFinitely2025}%
  \BibitemOpen
  \bibfield  {author} {\bibinfo {author} {\bibfnamefont {O.}~\bibnamefont {Ahmed}}, \bibinfo {author} {\bibfnamefont {F.}~\bibnamefont {Tennie}}, \ and\ \bibinfo {author} {\bibfnamefont {L.}~\bibnamefont {Magri}},\ }\href {\doibase 10.1007/s42484-025-00261-9} {\bibfield  {journal} {\bibinfo  {journal} {Quantum Machine Intelligence}\ }\textbf {\bibinfo {volume} {7}},\ \bibinfo {pages} {31} (\bibinfo {year} {2025}{\natexlab{a}})}\BibitemShut {NoStop}%
\bibitem [{\citenamefont {Wang}\ \emph {et~al.}(2024)\citenamefont {Wang}, \citenamefont {Sun},\ and\ \citenamefont {Zhang}}]{wang2024enhanced}%
  \BibitemOpen
  \bibfield  {author} {\bibinfo {author} {\bibfnamefont {L.}~\bibnamefont {Wang}}, \bibinfo {author} {\bibfnamefont {Y.}~\bibnamefont {Sun}}, \ and\ \bibinfo {author} {\bibfnamefont {X.}~\bibnamefont {Zhang}},\ }\href {\doibase 10.1103/PhysRevResearch.6.043183} {\bibfield  {journal} {\bibinfo  {journal} {Phys. Rev. Res.}\ }\textbf {\bibinfo {volume} {6}},\ \bibinfo {pages} {043183} (\bibinfo {year} {2024})}\BibitemShut {NoStop}%
\bibitem [{\citenamefont {Ahmed}\ \emph {et~al.}(2025{\natexlab{b}})\citenamefont {Ahmed}, \citenamefont {Tennie},\ and\ \citenamefont {Magri}}]{ahmed2025robust}%
  \BibitemOpen
  \bibfield  {author} {\bibinfo {author} {\bibfnamefont {O.}~\bibnamefont {Ahmed}}, \bibinfo {author} {\bibfnamefont {F.}~\bibnamefont {Tennie}}, \ and\ \bibinfo {author} {\bibfnamefont {L.}~\bibnamefont {Magri}},\ }\href@noop {} {\enquote {\bibinfo {title} {Robust quantum reservoir computers for forecasting chaotic dynamics: generalized synchronization and stability},}\ } (\bibinfo {year} {2025}{\natexlab{b}}),\ \Eprint {http://arxiv.org/abs/2506.22335} {arXiv:2506.22335 [quant-ph]} \BibitemShut {NoStop}%
\bibitem [{\citenamefont {Xu}\ \emph {et~al.}(2018)\citenamefont {Xu}, \citenamefont {Chen}, \citenamefont {Zeng}, \citenamefont {Zhang}, \citenamefont {Song}, \citenamefont {Liu}, \citenamefont {Guo}, \citenamefont {Zhang}, \citenamefont {Xu}, \citenamefont {Deng}, \citenamefont {Huang}, \citenamefont {Wang}, \citenamefont {Zhu}, \citenamefont {Zheng},\ and\ \citenamefont {Fan}}]{Xu_2018}%
  \BibitemOpen
  \bibfield  {author} {\bibinfo {author} {\bibfnamefont {K.}~\bibnamefont {Xu}}, \bibinfo {author} {\bibfnamefont {J.-J.}\ \bibnamefont {Chen}}, \bibinfo {author} {\bibfnamefont {Y.}~\bibnamefont {Zeng}}, \bibinfo {author} {\bibfnamefont {Y.-R.}\ \bibnamefont {Zhang}}, \bibinfo {author} {\bibfnamefont {C.}~\bibnamefont {Song}}, \bibinfo {author} {\bibfnamefont {W.}~\bibnamefont {Liu}}, \bibinfo {author} {\bibfnamefont {Q.}~\bibnamefont {Guo}}, \bibinfo {author} {\bibfnamefont {P.}~\bibnamefont {Zhang}}, \bibinfo {author} {\bibfnamefont {D.}~\bibnamefont {Xu}}, \bibinfo {author} {\bibfnamefont {H.}~\bibnamefont {Deng}}, \bibinfo {author} {\bibfnamefont {K.}~\bibnamefont {Huang}}, \bibinfo {author} {\bibfnamefont {H.}~\bibnamefont {Wang}}, \bibinfo {author} {\bibfnamefont {X.}~\bibnamefont {Zhu}}, \bibinfo {author} {\bibfnamefont {D.}~\bibnamefont {Zheng}}, \ and\ \bibinfo {author} {\bibfnamefont {H.}~\bibnamefont {Fan}},\ }\href {\doibase 10.1103/physrevlett.120.050507} {\bibfield  {journal} {\bibinfo
  {journal} {Physical Review Letters}\ }\textbf {\bibinfo {volume} {120}} (\bibinfo {year} {2018}),\ 10.1103/physrevlett.120.050507}\BibitemShut {NoStop}%
\bibitem [{\citenamefont {Huang}\ \emph {et~al.}(2020)\citenamefont {Huang}, \citenamefont {Wu}, \citenamefont {Fan},\ and\ \citenamefont {Zhu}}]{Huang_2020}%
  \BibitemOpen
  \bibfield  {author} {\bibinfo {author} {\bibfnamefont {H.-L.}\ \bibnamefont {Huang}}, \bibinfo {author} {\bibfnamefont {D.}~\bibnamefont {Wu}}, \bibinfo {author} {\bibfnamefont {D.}~\bibnamefont {Fan}}, \ and\ \bibinfo {author} {\bibfnamefont {X.}~\bibnamefont {Zhu}},\ }\href {\doibase 10.1007/s11432-020-2881-9} {\bibfield  {journal} {\bibinfo  {journal} {Science China Information Sciences}\ }\textbf {\bibinfo {volume} {63}} (\bibinfo {year} {2020}),\ 10.1007/s11432-020-2881-9}\BibitemShut {NoStop}%
\bibitem [{\citenamefont {Blais}\ \emph {et~al.}(2021)\citenamefont {Blais}, \citenamefont {Grimsmo}, \citenamefont {Girvin},\ and\ \citenamefont {Wallraff}}]{Blais_2021}%
  \BibitemOpen
  \bibfield  {author} {\bibinfo {author} {\bibfnamefont {A.}~\bibnamefont {Blais}}, \bibinfo {author} {\bibfnamefont {A.~L.}\ \bibnamefont {Grimsmo}}, \bibinfo {author} {\bibfnamefont {S.~M.}\ \bibnamefont {Girvin}}, \ and\ \bibinfo {author} {\bibfnamefont {A.}~\bibnamefont {Wallraff}},\ }\href {\doibase 10.1103/RevModPhys.93.025005} {\bibfield  {journal} {\bibinfo  {journal} {Rev. Mod. Phys.}\ }\textbf {\bibinfo {volume} {93}},\ \bibinfo {pages} {025005} (\bibinfo {year} {2021})}\BibitemShut {NoStop}%
\bibitem [{\citenamefont {Gong}\ \emph {et~al.}(2021)\citenamefont {Gong}, \citenamefont {de~Moraes~Neto}, \citenamefont {Zha}, \citenamefont {Wu}, \citenamefont {Rong}, \citenamefont {Ye}, \citenamefont {Li}, \citenamefont {Zhu}, \citenamefont {Wang}, \citenamefont {Zhao}, \citenamefont {Liang}, \citenamefont {Lin}, \citenamefont {Xu}, \citenamefont {Peng}, \citenamefont {Deng}, \citenamefont {Bayat}, \citenamefont {Zhu},\ and\ \citenamefont {Pan}}]{Gong2021Experimental}%
  \BibitemOpen
  \bibfield  {author} {\bibinfo {author} {\bibfnamefont {M.}~\bibnamefont {Gong}}, \bibinfo {author} {\bibfnamefont {G.~D.}\ \bibnamefont {de~Moraes~Neto}}, \bibinfo {author} {\bibfnamefont {C.}~\bibnamefont {Zha}}, \bibinfo {author} {\bibfnamefont {Y.}~\bibnamefont {Wu}}, \bibinfo {author} {\bibfnamefont {H.}~\bibnamefont {Rong}}, \bibinfo {author} {\bibfnamefont {Y.}~\bibnamefont {Ye}}, \bibinfo {author} {\bibfnamefont {S.}~\bibnamefont {Li}}, \bibinfo {author} {\bibfnamefont {Q.}~\bibnamefont {Zhu}}, \bibinfo {author} {\bibfnamefont {S.}~\bibnamefont {Wang}}, \bibinfo {author} {\bibfnamefont {Y.}~\bibnamefont {Zhao}}, \bibinfo {author} {\bibfnamefont {F.}~\bibnamefont {Liang}}, \bibinfo {author} {\bibfnamefont {J.}~\bibnamefont {Lin}}, \bibinfo {author} {\bibfnamefont {Y.}~\bibnamefont {Xu}}, \bibinfo {author} {\bibfnamefont {C.-Z.}\ \bibnamefont {Peng}}, \bibinfo {author} {\bibfnamefont {H.}~\bibnamefont {Deng}}, \bibinfo {author} {\bibfnamefont {A.}~\bibnamefont {Bayat}}, \bibinfo {author} {\bibfnamefont
  {X.}~\bibnamefont {Zhu}}, \ and\ \bibinfo {author} {\bibfnamefont {J.-W.}\ \bibnamefont {Pan}},\ }\href {\doibase 10.1103/PhysRevResearch.3.033043} {\bibfield  {journal} {\bibinfo  {journal} {Phys. Rev. Res.}\ }\textbf {\bibinfo {volume} {3}},\ \bibinfo {pages} {033043} (\bibinfo {year} {2021})}\BibitemShut {NoStop}%
\bibitem [{\citenamefont {Blatt}\ and\ \citenamefont {Roos}(2012)}]{blatt2012quantum}%
  \BibitemOpen
  \bibfield  {author} {\bibinfo {author} {\bibfnamefont {R.}~\bibnamefont {Blatt}}\ and\ \bibinfo {author} {\bibfnamefont {C.~F.}\ \bibnamefont {Roos}},\ }\href@noop {} {\bibfield  {journal} {\bibinfo  {journal} {Nature Physics}\ }\textbf {\bibinfo {volume} {8}},\ \bibinfo {pages} {277} (\bibinfo {year} {2012})}\BibitemShut {NoStop}%
\bibitem [{\citenamefont {Duan}\ \emph {et~al.}(2021)\citenamefont {Duan}, \citenamefont {Gong} \emph {et~al.}}]{monroe2021programmable}%
  \BibitemOpen
  \bibfield  {author} {\bibinfo {author} {\bibfnamefont {L.-M.}\ \bibnamefont {Duan}}, \bibinfo {author} {\bibfnamefont {Z.-X.}\ \bibnamefont {Gong}},  \emph {et~al.},\ }\href@noop {} {\bibfield  {journal} {\bibinfo  {journal} {Reviews of Modern Physics}\ }\textbf {\bibinfo {volume} {93}},\ \bibinfo {pages} {025001} (\bibinfo {year} {2021})}\BibitemShut {NoStop}%
\bibitem [{\citenamefont {Katz}\ \emph {et~al.}(2023)\citenamefont {Katz}, \citenamefont {Cetina},\ and\ \citenamefont {Monroe}}]{katz2023programmable}%
  \BibitemOpen
  \bibfield  {author} {\bibinfo {author} {\bibfnamefont {O.}~\bibnamefont {Katz}}, \bibinfo {author} {\bibfnamefont {M.}~\bibnamefont {Cetina}}, \ and\ \bibinfo {author} {\bibfnamefont {C.}~\bibnamefont {Monroe}},\ }\href {\doibase 10.1103/PRXQuantum.4.030311} {\bibfield  {journal} {\bibinfo  {journal} {PRX Quantum}\ }\textbf {\bibinfo {volume} {4}},\ \bibinfo {pages} {030311} (\bibinfo {year} {2023})}\BibitemShut {NoStop}%
\bibitem [{\citenamefont {Noh}\ and\ \citenamefont {Angelakis}(2016)}]{noh2016quantum}%
  \BibitemOpen
  \bibfield  {author} {\bibinfo {author} {\bibfnamefont {C.}~\bibnamefont {Noh}}\ and\ \bibinfo {author} {\bibfnamefont {D.~G.}\ \bibnamefont {Angelakis}},\ }\href {\doibase 10.1088/0034-4885/80/1/016401} {\bibfield  {journal} {\bibinfo  {journal} {Rep. Prog. Phys.}\ }\textbf {\bibinfo {volume} {80}},\ \bibinfo {pages} {016401} (\bibinfo {year} {2016})}\BibitemShut {NoStop}%
\bibitem [{\citenamefont {Wang}\ \emph {et~al.}(2019)\citenamefont {Wang}, \citenamefont {Qiu}, \citenamefont {Xiao}, \citenamefont {Zhan}, \citenamefont {Bian}, \citenamefont {Yi},\ and\ \citenamefont {Xue}}]{wang2019simulating}%
  \BibitemOpen
  \bibfield  {author} {\bibinfo {author} {\bibfnamefont {K.}~\bibnamefont {Wang}}, \bibinfo {author} {\bibfnamefont {X.}~\bibnamefont {Qiu}}, \bibinfo {author} {\bibfnamefont {L.}~\bibnamefont {Xiao}}, \bibinfo {author} {\bibfnamefont {X.}~\bibnamefont {Zhan}}, \bibinfo {author} {\bibfnamefont {Z.}~\bibnamefont {Bian}}, \bibinfo {author} {\bibfnamefont {W.}~\bibnamefont {Yi}}, \ and\ \bibinfo {author} {\bibfnamefont {P.}~\bibnamefont {Xue}},\ }\href {\doibase 10.1103/PhysRevLett.122.020501} {\bibfield  {journal} {\bibinfo  {journal} {Phys. Rev. Lett.}\ }\textbf {\bibinfo {volume} {122}},\ \bibinfo {pages} {020501} (\bibinfo {year} {2019})}\BibitemShut {NoStop}%
\end{thebibliography}%

\clearpage
\onecolumngrid
% \beginsupplement
% \section{Supplemental Material}

%%%%%%%%%%%%%%%%%%%%%%%%%%%%%%%%%%%%%%%%%%%%%%%%%%%%%%%%%%%%%%%%%%%%%%%%%%%%
%%%%%%%%%%%%%%%%%%%%%%%%%%%%%%%  ABSTRACT  %%%%%%%%%%%%%%%%%%%%%%%%%%%%%%%%%
%%%%%%%%%%%%%%%%%%%%%%%%%%%%%%%%%%%%%%%%%%%%%%%%%%%%%%%%%%%%%%%%%%%%%%%%%%%%

%\tableofcontents

%%%%%%%%%%%%%%%%%%%%%%%%%%%%%%%%%%%%%%%%%%%%%%%%%%%%%%%%%%%%%%%%%%%%%%%%%%%%
%%%%%%%%%%%%%%%%%%%%%%%%%%%%%%%%%%%%%%%%%%%%%%%%%%%%%%%%%%%%%%%%%%%%%%%%%%%%
%%%%%%%%%%%%%%%%%%%%%%%%%%%%%%%%%%%%%%%%%%%%%%%%%%%%%%%%%%%%%%%%%%%%%%%%%%%%

\end{document}